\documentclass[useAMS,usenatbib,onecolumn]{mn2e}
\usepackage{graphicx}
\usepackage{amsmath,amsfonts,amssymb}
\usepackage{wrapfig}
\usepackage{epsfig}
\usepackage{psfrag}
\usepackage{subfigure}
\usepackage{pstricks,pst-plot} 

\begin{document}
%%%%%%%%%%%%%%%%%%%%%%%%%%%%%%%%%%%%%%%%%%%%%%%%

\title[Detectability of eccentric binary pulsars]
{On the detectability of eccentric binary pulsars}

\author[Bagchi, Lorimer \& Wolfe]
{\parbox[t]{\textwidth}{Manjari Bagchi$^{1}$\thanks{Email: Manjari.Bagchi@mail.wvu.edu}, Duncan R. Lorimer$^{1,2,3}$ and Spencer Wolfe$^1$ }\\
\vspace*{3pt} \\
\\ $^1$ Department of Physics, White Hall, West Virginia University, Morgantown, WV 26506, USA
\\ $^2$ NRAO, Green Bank Observatory, PO Box 2, Green Bank, WV 24944, USA
\\$^3$ Astrophysics, University of Oxford, Denys Wilkinson Building, Keble Road, Oxford OX1 3RH
}

\maketitle

\begin{abstract}
By generalizing earlier work of Johnston \& Kulkarni, we present
a detailed description of the reduction in the signal-to-noise ratio
for observations of binary pulsars. We present analytical 
expressions, and provide software, to calculate the sensitivity reduction for orbits
of arbitrary eccentricity. We find that this reduction can be quite significant, especially in the case of a massive companion like another neutron star or a black hole. On the other hand, the reduction is less for highly eccentric orbits. We also demonstrate that this loss of sensitivity can be recovered by employing ``acceleration search" or ``acceleration-jerk search" algorithms.
\end{abstract}


\section{Introduction} 
\label{lb:introductions} 

It is now well established that studies of binary pulsar systems
provide a unique means to test relativistic theories of gravity in the
strong-field regime \citep{stairs03, ks08}, with the most stringent
tests often possible in systems with short orbital periods and high
eccentricities. Detecting such systems in pulsar searches is
challenging due to the Doppler modulation of the pulsed signal as the
pulsar moves with respect to the centre of mass
of the binary system during an observation. In an
earlier study, Johnston \& Kulkarni (1991; \nocite{jk91} hereafter
JK91) derived an analytical framework to compute the
reduction of the signal-to-noise ratio due to binary motion for the
case of circular orbits. This work also considered the improvement in
sensitivity due to ``acceleration searches'' in which the data are
corrected for binary motion either in the time \citep{and93} or
frequency domain \citep{ran01}.

Although circular orbits are applicable to many known binary pulsar systems \citep{lor08}, for the purposes of a growing number of studies,
it is desirable to compute the loss of sensitivity in eccentric binary
systems for different types of companions.  The purpose of this paper
is to extend the work of JK91 to consider the case of orbits with
arbitrary eccentricity. This framework allows us to compute the
sensitivity degradation for any type of binary system, therefore
allowing detailed investigation of survey sensitivity limits, which is
applicable to a number of statistical studies (see \citet{rl10} for example). While \citet{rzw98}
investigated this, they did not provide explicit analytical
expressions, nor did they extensively explore the parameter space
implied by pulsar surveys and the different types of binary systems. We
will discuss more about their work later when appropriate.

The plan for the rest of this paper is as follows. The analytical
expressions to describe this problem are developed in Section
\ref{sec:form}, numerical methods to employ those expressions are
elaborated in Section \ref{sec:analysis}, and some applications of the
model in different cases are reported in Section \ref{sec:res}.
Finally, our conclusions are given in Section \ref{sec:conclu}.

\section{Formulation}
\label{sec:form} 

Following JK91, we consider a pulsar with spin period $P_{p}$. The
signal emitted from the pulsar at an arbitrary time $t$ can be written
as a summation of Fourier components:
\begin{equation}
S_{p}(t) = \sum_{k=1}^{k=\infty}\, a_{k} \, {\rm exp}\left(ik \omega_{p} t + i \psi_{k} \right),
\label{eq:sig_puls}
\end{equation} 
where $\omega_{p}= 2 \pi / P_{p} $ is the angular spin frequency of
the pulsar, $\psi_{k}$ is the phase factor of the $k^{\rm th}$ component
and $i=\sqrt{-1}$. The signal received on the Earth
\begin{equation}
S_{R}(t) = \sum_{k=1}^{k=\infty}\, a_{k} \, {\rm exp} \left(ik \omega_{p} \left(t + \frac{d}{c} \right) + i \psi_{k} \right),
\label{eq:sig_obs}
\end{equation} 
where $d$ is the distance between the pulsar and the Earth at time $t$ 
and $c$ is the speed of light.
The $m^{\rm th}$ harmonic of the received signal is
\begin{equation}
S_{R}^{m}(t) =  a_{m} \, {\rm exp} \left(im \omega_{p} \left(t + \frac{d}{c} \right) + i \psi_{m} \right).
\label{eq:sigm_obs}
\end{equation} 
Using a Taylor series expansion of $d$, we can write the distance to the pulsar
\begin{equation}
d = d_{0} + v_{l0} t + \frac{a_{l0} t^2}{2!} + \frac{j_{l0} t^3}{3!} + \ldots,
 \label{eq:d_taylor}
\end{equation} 
where $v_{l}$ is the line-of-sight velocity, $a_{l}$ is the
line-of-sight acceleration, and $j_{l}$ is the line-of-sight jerk at
the time $t$; $v_{l0}$, $a_{l0}$ and $j_{l0}$ are the values of these
quantities at $t=0$. $d_{0}$ is the distance between the pulsar and
the Earth at $t=0$. An isolated pulsar moves with a constant velocity,
so $a_{l}$, $j_{l}$ (and higher order terms) are zero; but, this is
not the case for a binary pulsar. Similarly, for $v_l$, the Taylor
series expansion gives
\begin{equation}
v_{l} = v_{l0}  + a_{l0} t+ \frac{j_{l0} t^2}{2!} + \ldots
\label{eq:v_taylor}
\end{equation}
Integrating this expression with respect to time, we get
\begin{equation}
\int_{0}^{t} v_{l} \, dt = \int_{0}^{t} \left( v_{l0}  + a_{l0} t+ \frac{j_{l0} t^2}{2!} + \ldots \right) \, dt = v_{l0} t + \frac{a_{l0} t^2}{2!} + \frac{j_{l0} t^3}{3!} + \ldots
\label{eq:v_taylorint}
\end{equation} 
So Equation (\ref{eq:d_taylor}) becomes
\begin{equation}
d = d_{0} + \int_{0}^{t} v_{l} \, dt.
 \label{eq:d_exp1}
\end{equation} 
To detect a pulsar, one performs a Fourier transform of the time
series of the received signal as $\mathcal{F} \left[ S_{R}(t) \right]
= \int_{0}^{T} S_{R}(t) \, {\rm exp} (-i \omega t) \, dt$ which gives
the power as $ \bigr \rvert \mathcal{F} \left[ S_{R}(t) \right] \bigl
\lvert ^2 $. Here $T$ is the duration of the observation. Because the
signal has a small duty cycle instead of being purely sinusoidal,
several harmonics are present in the Fourier power spectrum. One
usually performs ``harmonic summing'' up to the $n^{\rm th}$ harmonic to
increase the search sensitivity by a factor of up to $\sqrt{n}$
\nocite{lk05} (see, e.g., Lorimer \& Kramer 2005). From
Equation (\ref{eq:sigm_obs}), the power of the $m^{\rm th}$ harmonic in the
Fourier spectrum can be written as 
\begin{eqnarray}
 \bigl \lvert \mathcal{F} \left[ S_{R}^{m}(t) \right] \bigr \rvert ^2  =   \Bigl \lvert \int_{0}^{T} \, a_{m} \, {\rm exp} \left(im \omega_{p} \left(t + \frac{d}{c} \right) + i \psi_{m} \right) \, {\rm exp} (-i \omega t) \, dt \Bigr \rvert ^2  \nonumber \\  = \Bigl \lvert \int_{0}^{T} \, a_{m}  {\rm exp} \left(im \omega_{p} \left(t + \frac{1}{c} \, (d_{0} + v_{l0} t + \frac{a_{l0} t^2}{2!} + \frac{j_{l0} t^3}{3!} + \ldots) \right) + i \psi_{m} \right)  {\rm exp} (-i \omega t) \, dt \Bigr \rvert ^2 \nonumber \\  = \Bigl \lvert \int_{0}^{T} \, a_{m} {\rm exp} \left( \frac{im \omega_{p} d_{0}}{c} + i \psi_{m}  \right)  {\rm exp} \left( \frac{im \omega_{p}}{c} \left[ (1+\frac{v_{l0}}{c}) ct   + \, \frac{a_{l0} t^2}{2!} + \frac{j_{l0} t^3}{3!} + \ldots \right]  \right)  {\rm exp} (-i \omega t) \, dt \Bigr \rvert ^2,
\label{eq:powm_fourier}
\end{eqnarray} where we note that $d_{0}$ contributes only as a constant phase factor and 
\begin{equation}
m \, \omega_{p}^{'} = m \, \omega_{p} \left[ (1+\frac{v_{l0}}{c}) + \, \frac{a_{l0} t}{c \, 2!} + \frac{j_{l0} t^2}{c \, 3!} + \ldots \right] 
\label{eq:ob_angfreq}
\end{equation} 
is the observed angular spin frequency of the $m^{\rm th}$ harmonic. As $m
\, \omega_{p}^{'}$ changes with $t$, the power gets distributed over
adjacent Fourier bins leading to a loss in sensitivity. Nevertheless,
the pulsar can be detected in the Fourier bin where the power is
maximum. In the case of a pulsar moving with a constant velocity, the
power of the $m^{\rm th}$ harmonic will be peaked at the Fourier bin of
angular frequency $m \, \omega_{p} \, (1+\frac{v_{l0}}{c})$ which is
the conventional Doppler effect.

Following JK91 let us define three efficiency factors,
$\gamma_{1m}$, $\gamma_{2m}$ and $\gamma_{3m}$. The factor $\gamma_{1m}^2$ is
the ratio of the height of the power of the $m^{\rm th}$ harmonic in the
Fourier spectrum when the acceleration, jerk, and other higher order
derivatives of the pulsar are non-zero, to the height when these terms
are zero, \textit{i.e.}, the pulsar has a constant velocity. This can
be written as follows 
\begin{eqnarray}
\gamma_{1m}(\alpha_v, T)  =  \frac{1}{T}  \Bigl\lvert  \int_{0}^{T} {\rm exp} \left[ \frac{i m \omega_p}{c} \left( v_{l0} t + \frac{a_{l0} t^2}{2!} + \frac{j_{l0} t^3}{3!} + \ldots - \alpha_v t \right) \right] dt  \Bigr\rvert = \frac{1}{T}  \Bigl\lvert  \int_{0}^{T} {\rm exp} \left[ \frac{i m \omega_p}{c} \left( \left( \int_{0}^{t} v_{l} \, dt  \right)- \alpha_v t \right) \right] dt  \Bigr\rvert,
\label{eq:gamma1_velocity}
\end{eqnarray} 
where $\alpha_v$ is a free parameter. The pulsar will be detected for
such a value of $\alpha_v$ which maximizes $\gamma_{1m}$ with apparent
frequency of the $m^{\rm th}$ harmonic as $m \omega_{p}^{'} = m \omega_{p}
\left( 1 + \alpha_{v}/c \right)$. Clearly, the maximum value of
$\gamma_{1m}^2 = 1$ is possible only when the pulsar moves with a
constant velocity, giving $\alpha_v = v_{l0}$.

The above efficiency factor $\gamma_{1m}$ essentially describes the
sensitivity loss of a standard pulsar search.  Acceleration searches
attempt to improve upon this efficiency by accounting for the Doppler
shifting of the pulsar signal during the observation. A review of some
of the various techniques that have been developed so far can be found
in \citet{lk05}. The most common type of acceleration
search assumes that the line-of-sight acceleration during the
observation is a constant value. This is a good approximation for
binary periods that are significantly longer than the survey
integration time, and we consider these searches using the efficiency
factor $\gamma_{2m}$.  Specifically, $\gamma_{2m}^2$ is the ratio of
the height of the power of the $m^{\rm th}$ harmonic in the Fourier
spectrum when the jerk and other higher order derivatives of the
pulsar are non-zero, to the height when these terms are zero, \textit{i.e.}, the pulsar has a constant acceleration. Expressing this idea mathematically,
we have 
\begin{eqnarray}
\gamma_{2m}(\alpha_a, \alpha_v, T)  =  \frac{1}{T} \Bigl\lvert  \int_{0}^{T} {\rm exp} \left[ \frac{i m \omega_p}{c}   \left( v_{l0} t + \frac{a_{l0} t^2}{2!} + \frac{j_{l0} t^3}{3!} + \ldots  - \alpha_a t^2 -\alpha_v t \right) \right] dt  \Bigr\rvert  \nonumber \\  =  \frac{1}{T}  \Bigl\lvert  \int_{0}^{T} {\rm exp} \left[ \frac{i m \omega_p}{c} \left( \left( \int_{0}^{t} v_{l} \, dt  \right) - \alpha_a t^2 - \alpha_v t \right) \right] dt  \Bigr\rvert,
\label{eq:gamma2_velocity}
\end{eqnarray} 
where $\alpha_a$, $\alpha_v$ are free parameters. The use of an acceleration search algorithm leads to the detection of the pulsar for such a set
of values of $\alpha_a$ and $\alpha_v$ which maximizes
$\gamma_{2m}$. Clearly, the maximum value of $\gamma_{2m}^2 = 1$ is
possible only when the pulsar moves with a constant acceleration,
giving $\alpha_v = v_{l0}$ and $\alpha_a = a_{l0}/2!$.

Due to the additional computational requirements, searches involving
the unknown jerk term are less commonly carried out. To investigate
their relative efficiency, however, 
we define $\gamma_{3m}$ such that $\gamma_{3m}^2$ is the ratio of the
height of the power of the $m^{\rm th}$ harmonic in the Fourier spectrum
when the derivative of the jerk and other higher order derivatives of
the pulsar are non-zero, to the height when these terms are zero, \textit{i.e.}, the pulsar has a constant jerk. With this in mind, we may write
\begin{eqnarray}
\gamma_{3m}(\alpha_j, \alpha_a, \alpha_v, T)  =  \frac{1}{T} 
 \Bigl\lvert  \int_{0}^{T} {\rm exp} \left[ \frac{i m \omega_p}{c}  \left(  v_{l0} t + \frac{a_{l0} t^2}{2!} + \frac{j_{l0} t^3}{3!} + \ldots  - \alpha_j t^3 - \alpha_a t^2 -\alpha_v t \right) \right] dt  \Bigr\rvert  \nonumber \\ = \frac{1}{T}  \Bigl\lvert  \int_{0}^{T} {\rm exp} \left[ \frac{i m \omega_p}{c} \left( \left( \int_{0}^{t} v_{l} \, dt \right) - \alpha_j t^3 - \alpha_a t^2 - \alpha_v t \right) \right] dt  \Bigr\rvert,
\label{eq:gamma3_velocity}
\end{eqnarray} 
where $\alpha_j$, $\alpha_a$, and $\alpha_v$ are free parameters. 
Such a search process would detect a pulsar with 
a set of values of $\alpha_j$, $\alpha_a$ and
$\alpha_v$ which maximizes $\gamma_{3m}$. Clearly, the maximum value
of $\gamma_{3m}^2 = 1$ is possible only when the pulsar moves with a
constant jerk, giving $\alpha_v = v_{l0}$, $\alpha_a = a_{l0}/2!$ and
$\alpha_j = j_{l0}/3!$.

\begin{figure}
\centerline{\psfig{figure=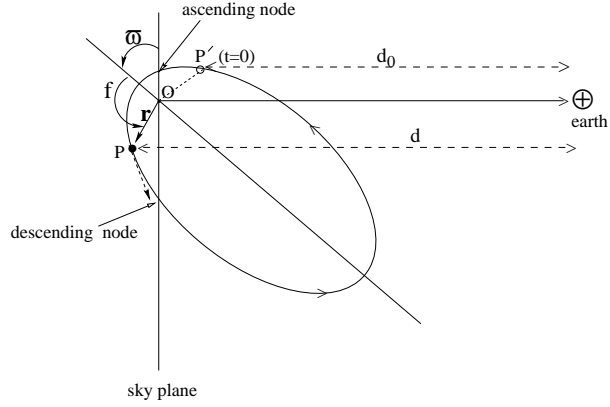,width=8cm,angle=0}}
\caption{The orbit of a pulsar in a binary system, projected onto a
  plane containing the line of sight, \textit{i.e.}, perpendicular to
  the sky plane. The semi-major axis of the projected ellipse
  $a_{p}^{'} = a_{p} \, {\rm sin}\,i$ where $i$ is the orbital
  inclination angle and $a_{p}$ is the semi-major axis of the actual
  orbit of the pulsar. $P^{'}$ is the location of the pulsar at $t=0$
  and $P$ is the location of the pulsar at any arbitrary time $t$. $f$
  is the true anomaly and $\varpi$ is the longitude of the
  periastron. $d_{0}$ is the distance between the pulsar and the Earth
  at $t=0$ and $d$ is the distance between the pulsar and the Earth at
  $t$.} \label{fig:orbitgeometry}
\end{figure}

To compute the relevant line-of-sight terms in the above expressions,
we need to consider the shape of the pulsar orbit.  We denote 
the semi-major axis of the pulsar orbit to be $a_{p}$ and the semi-major
axis of the ellipse projected perpendicular to the plane of the sky to be
$a_{p}^{'} = a_{p} \, {\rm sin}\,i$ where $i$ is the orbital
inclination angle. This projected ellipse is shown in
Fig.~\ref{fig:orbitgeometry} where $f$ is the true anomaly and
$\varpi$ is the longitude of the periastron. The pulsar is located at
$P^{'}$ at $t=0$ and at $P$ at an arbitrary time $t$. If ${\bf r}$ is
the radius vector of the position of the pulsar at $t$ with magnitude
$r$, then the projection of $r$ along the line of sight is given by
\begin{equation}
r_{l} = r \,  {\rm sin}\,(f+ \varpi ).
\label{eq:r_rl}
\end{equation} 
Similarly, the projection of the magnitude $r_{0}$ of the radius 
vector of the pulsar at $t=0$ can be written as
\begin{equation}
r_{l0} = r_{0} \,  {\rm sin}\,(f_{0} + \varpi),  
\label{eq:r_rl0}
\end{equation} 
where $f_{0}$ is the true anomaly at $t=0$. We can write
\begin{equation}
\int_{0}^{t} v_{l} \, dt = ( {r_{l} - r_{l0}} ).
\label{eq:vltor_rl}
\end{equation} So, Equation (\ref{eq:d_exp1}) becomes
\begin{equation}
d = d_{0} + \left( {r_{l} - r_{l0}} \right).
\label{eq:d_exp2}
\end{equation} 

Using Equation (\ref{eq:vltor_rl}), Equations
(\ref{eq:gamma1_velocity}), (\ref{eq:gamma2_velocity}) and
(\ref{eq:gamma3_velocity}) become
\begin{equation}
\gamma_{1m}(\alpha_v, T)  =  \frac{1}{T}  \Bigl\lvert  \int_{0}^{T} {\rm exp} \left[ \frac{i m \omega_p}{c} \left( r_{l} - r_{l0} - \alpha_v t \right) \right] dt  \Bigr\rvert,
\label{eq:gamma1_velocitysimple}
\end{equation} 
\begin{equation}
\gamma_{2m}(\alpha_a, \alpha_v, T)  =  \frac{1}{T}  \Bigl\lvert  \int_{0}^{T} {\rm exp} \left[ \frac{i m \omega_p}{c} \left( r_{l} - r_{l0} - \alpha_a t^2- \alpha_v t \right) \right] dt  \Bigr\rvert   
\label{eq:gamma2_velocitysimple}
\end{equation} 
and
\begin{equation}
\gamma_{3m}(\alpha_j, \alpha_a, \alpha_v, T) =  \frac{1}{T} \Bigl\lvert  \int_{0}^{T} {\rm exp} \left[ \frac{i m \omega_p}{c} \left( r_{l} - r_{l0} - \alpha_j t^3 - \alpha_a t^2- \alpha_v t \right) \right] dt  \Bigr\rvert.
\label{eq:gamma3_velocitysimple}
\end{equation} 

For the general case of an orbit of eccentricity $e$, we have
\begin{equation}
r = a_{p}^{'} \left(1 - e^2 \right) \, (1 + e \,{\rm cos}\,f )^{-1}
\label{eq:r_ellipse}
\end{equation} 
so that
\begin{equation}
r_{0} = a_{p}^{'} \left(1 - e^2 \right) \, (1 + e \,{\rm cos}\,f_{0} )^{-1}.
\label{eq:r0_ellipse}
\end{equation}
With this definition, Equations (\ref{eq:r_rl}) and (\ref{eq:r_rl0}) become
\begin{equation}
r_{l} = a_{p}^{'} \left(1 - e^2 \right) \, (1 + e \,{\rm cos}\,f )^{-1}\,  {\rm sin}\,(f + \varpi),
\label{eq:r_ellipse2}
\end{equation} and
\begin{equation}
r_{l0} = a_{p}^{'} \left(1 - e^2 \right) \, (1 + e \,{\rm cos}\,f_{0} )^{-1}\,  {\rm sin}\,(f_{0} + \varpi).
\label{eq:r0_ellipse2}
\end{equation}
Now, differentiating Equation (\ref{eq:r_rl}) and using Equation
(\ref{eq:r_ellipse}) we get the expression for the velocity along the
line of sight as
\begin{equation}
v_{l} = \dot{r_{l}} = \frac{2 \pi}{ P_{o}} \, \frac{a_{p}^{'}}{\sqrt{1-e^2}} \, \left[ {\rm cos}\,(f+ \varpi) + e \, {\rm cos}\,(\varpi) \right],
\label{eq:vl}
\end{equation} 
where $P_{o}$ is the orbital period of the pulsar and the orbital
angular frequency is $\omega_{o}= 2 \pi / P_{o} $. The
quantities $a_{p}^{'}$ and $P_{o}$ are related by Kepler's third law as
\begin{equation}
a_{p}^{'} = a_{p} \, {\rm sin} \, i 
= a_{R} \, \frac{M_{c}}{M_{p} + M_{c}} \, {\rm sin} \, i = \left[ \left( \frac{P_{o}}{2 \pi} \right)^2 \,  G \left( M_{p} + M_{c} \right) \right]^{1/3} \, \frac{M_{c}}{M_{p} + M_{c}} \, {\rm sin} \, i,
\label{eq:keplerapPo}
\end{equation} 
where $M_{p}$ is the mass of the pulsar and $M_{c}$ is the mass of the 
companion and $G$ is Newton's gravitational constant.

Differentiating Equation (\ref{eq:vl}), we get the expression for the
line-of-sight acceleration
\begin{equation}
a_{l} = \dot{v_{l}} = - \left( \frac{2 \pi}{ P_{o}} \right)^{2} \, \frac{a_{p}^{'}}{ \left(1-e^2 \right)^{2} } \, {\rm sin}\,(f+ \varpi) \, (1 + e \,{\rm cos}\,f )^{2}.
\label{eq:al}
\end{equation}
Similarly, differentiating Equation (\ref{eq:al}), 
we get the expression for the 
line-of-sight jerk
\begin{equation}
j_{l}   = - \left( \frac{2 \pi}{ P_{o}} \right)^{3} \, \frac{a_{p}^{'}}{ \left(1-e^2 \right)^{7/2} } \, (1 + e \,{\rm cos}\,f )^{3}   \left[ {\rm cos}\,(f+ \varpi) + e \, {\rm cos}\,(\varpi) - 3 e \, {\rm sin}\,(f+ \varpi ) \, {\rm sin}\,(f) \right].
\label{eq:jl}
\end{equation}
In principle, one can continue differentiating to get higher order
derivatives (e.g. the ``jounce'' is the time derivative of the jerk).
The expressions for $v_{l}$ and $a_{l}$ were previously
derived by \citet{fkl01,fkl09}. Assigning $f=f_{0}$ in
Equations (\ref{eq:vl}), (\ref{eq:al}) and (\ref{eq:jl}) we get the values
of $v_{l0}$, $a_{l0}$ and $j_{l0}$ respectively.

To compute Equations (\ref{eq:gamma1_velocitysimple}),
(\ref{eq:gamma2_velocitysimple}) and (\ref{eq:gamma3_velocitysimple})
numerically, we need to solve Kepler's equations given below
\begin{subequations}
\begin{equation}
M = \omega_{o} (t - T_{p} )
\label{eq:kepler1}
\end{equation}
\begin{equation}
E - e \, {\rm sin} \, E = M
\label{eq:kepler2}
\end{equation}
\begin{equation}
f = 2 \, {\rm tan }^{-1} \left[ \sqrt{\frac{1+e}{1-e}} \, {\rm tan} \, \frac{E}{2}   \right] ,
\label{eq:kepler3}
\end{equation}
\end{subequations} 
where $M$ is the mean anomaly, $E$ is the eccentric anomaly and
$T_{p}$ is the epoch of the periastron passage. $M_{0} = - \omega_{o}
T_{p}$, $M_{0}$ being the mean anomaly at $t=0$.  For a circular
otbit, $e =0$, $\varpi = 0$, $f = E = M$, giving $f = \omega_{o} t +
f_{0}$, $f_{0} = M_{0} = - \omega_{o} T_{p} $ and we can get the
expressions given by JK91 by substituting $A = a_{p}^{'}
\omega_{o} $ and $\phi = f_{0} + \pi/2 $ (or if $\varpi =
270^{\circ}$, $\phi = f_{0}$).

\begin{figure}
\centerline{\psfig{figure=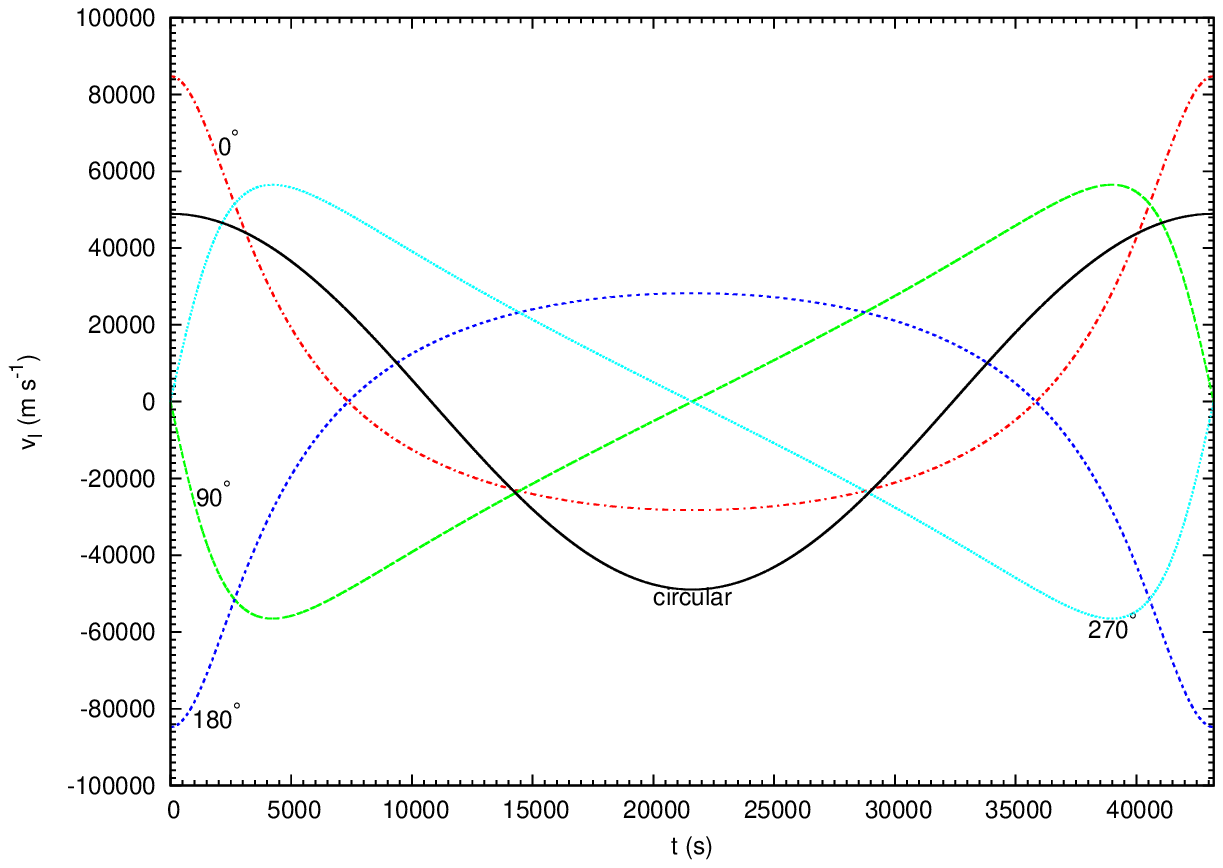,width=8cm,angle=0}}
\centerline{\psfig{figure=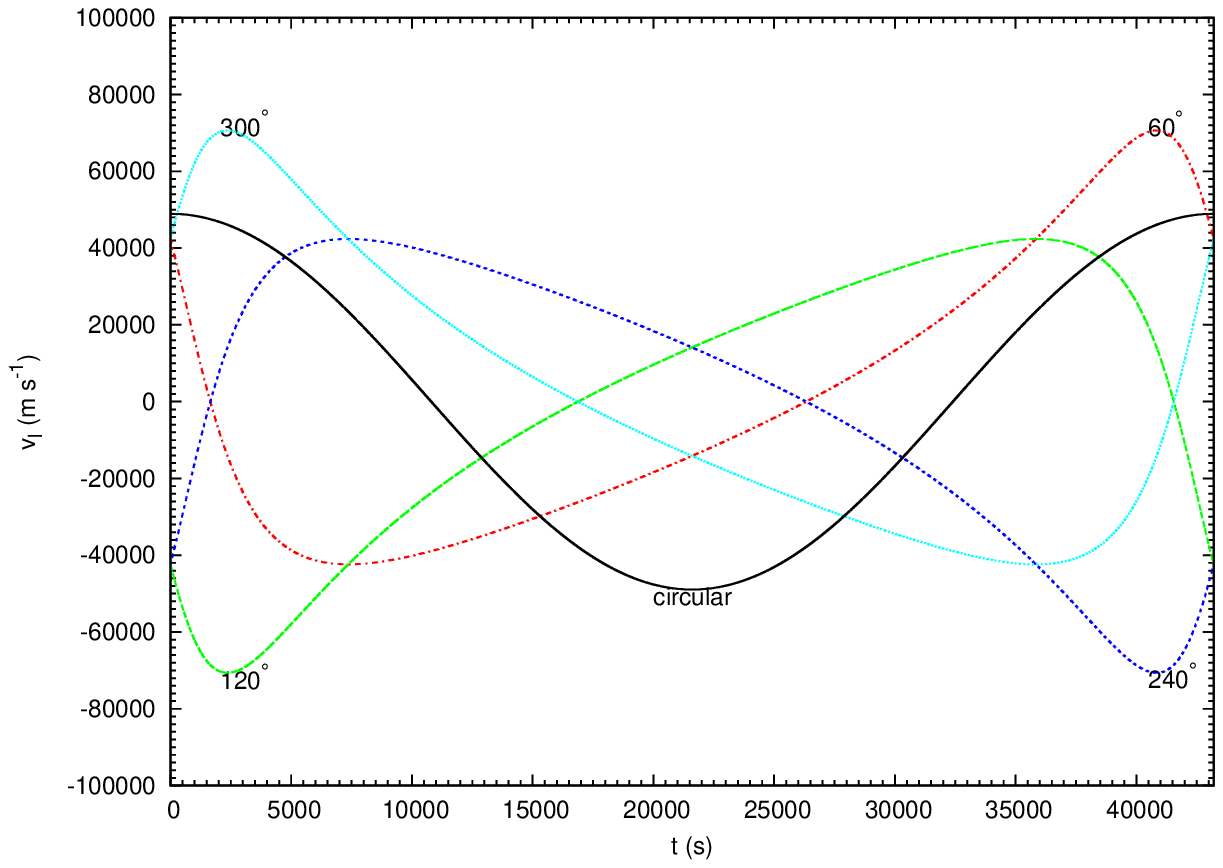,width=8cm,angle=0}}
\caption{Variation of the line-of-sight velocity with time over a complete
  orbit for a binary pulsar with $M_{p}=1.4~{M_{\odot}}$,
  $M_{c}=0.3~{M_{\odot}}$, $P_{o}=0.5$ day, $i = 60^{\circ}$, $T_p =
  0$, $e=0.5$ for different $\varpi$, and the line-of-sight velocity for the 
  same binary having zero eccentricity. 
} \label{fig:radialvelnswd}
\end{figure}

In Fig.~\ref{fig:radialvelnswd}, we plot the line-of-sight velocity with
respect to time over a complete orbit for a binary pulsar with
$M_{p}=1.4~{M_{\odot}}$, $M_{c}=0.3~{M_{\odot}}$, $P_{o}=0.5$ day, $i
= 60^{\circ}$, $T_p = 0$, $e=0.5$ for different $\varpi$ between
$0^{\circ}-360^{\circ}$ and compare with the line-of-sight velocity plot in
case of zero eccentricity. Like the circular case, line-of-sight velocity
curve for the eccentric orbit is also symmetric over the half orbital
period when $\varpi$ is either $0^{\circ}$, $90^{\circ}$,
$180^{\circ}$ or $270^{\circ}$ (upper panel), but not for any other
values of $\varpi$ (lower panel). In the lower panel of
Fig. \ref{fig:radialvelnswd}, $\varpi$ is chosen to be $60^{\circ}$,
$120^{\circ}$, $240^{\circ}$ and $300^{\circ}$, each in a different
quadrant, but have the same average line-of-sight velocity over the entire
orbit. The same can be found for other values of $\varpi$. That is
why, in the next section, when we perform our analysis for different
$\varpi$, we choose values only in the ranges of $0^{\circ} - 90^{\circ}$. The high value of eccentricity chosen here is justifiable
as binary pulsars with eccentricity 0.5 or even higher exist in globular clusters, mainly as a result of stellar encounters
(for a review, see Camilo \& Rasio 2004). \nocite{cr04}

For relativistic binaries, $P_{o}$, $e$, and $\varpi$ change with
time, and the rates of change of these parameters are classified as
Post-Keplarian parameters. These changes need to be considered
while performing pulsar timing analysis which involves pulse arrival
times over a long span of time (days or even years); but, for the
present study, we are concerned about the observability of pulsars,
and these changes are negligible over the duration of any particular
observation.

\section{Analysis}
\label{sec:analysis}

To demonstrate the impact of orbital motion on detectability, we 
will use the above framework to determine the maximum value of
$\gamma_{1m}$, $\gamma_{2m}$ and $\gamma_{3m}$. As different stellar
and orbital parameters like $P_{p}$, $P_{o}$, $M_{c}$, $M_{p}$, ${\rm
  sin} \, i$, $\varpi$ come into the equations given in Section
\ref{sec:form}, we perform our analysis for different values of these
parameters. For a particular binary, the above parameters are fixed as the
change in these parameters can be neglected over the duration of the
observation. The orbital phase, i.e. the true anomaly $f$, is, however, not known a priori, and changes during the observation.  

We divide the full-orbit (because of the asymmetry of an eccentric orbit which we 
have mentioned earlier) in equal steps of $10^{\circ}$, having
$f_{0,q}$s as $0^{\circ}, 10^{\circ}, 20^{\circ} \ldots ~ 350$ ($q$ going from 0 to 36), where $f_{0,q}$s are the values of the true anomaly at the start of the observation. For each $f_{0,q}$, we calculate $E_{0,q}$, $M_{0,q}$ and $T_{p,q}$ using
Kepler's equations (\ref{eq:kepler1}, \ref{eq:kepler2},
\ref{eq:kepler3}). We use these $T_{p,q}$s to calculate $M$, $E$ and $f$
at any time $t$ (between $0$ and $T$) again solving equations (\ref{eq:kepler1},
\ref{eq:kepler2}, \ref{eq:kepler3}). This takes care of the change in orbital phase during the observation. Then we use Equations (\ref{eq:vl}), (\ref{eq:al}) and (\ref{eq:jl}) to get $\alpha_v = v_{l}(t)$,
$\alpha_a = a_{l}(t)/2!$ and $\alpha_j = j_{l}(t)/3!$ for different
choices of $t$ between $0$ and $T$ and, for each set of $\alpha_v$,
$\alpha_a$ and $\alpha_j$, we perform the integrations given in
Equations (\ref{eq:gamma1_velocitysimple}), (\ref{eq:gamma2_velocitysimple}) and (\ref{eq:gamma3_velocitysimple})
numerically. The maximum values of of these integrations are the desired values $\gamma_{m1q}$, $\gamma_{m2q}$ and $\gamma_{m3q}$
respectively\footnote{Remember that only $\alpha_{v}$ comes in the expression for $\gamma_{m1q}$, and only $\alpha_{a}$ and
$\alpha_{v}$ come in the expression of $\gamma_{m2q}$.}. After this stage, we perform the weighted average of these efficiency factors over different $f_{0,q}$. 

To perform this weighted average, we need to calculate the probability of each $f_{0,q}$ using the fact that the probability of a pulsar to be at a particular position in the orbit is directly proportional to the time it spends at that position. As the areal
velocity is constant (Kepler's second law), the ratio of the time spent at true anomalies
$f_{1}$ and $f_{2}$ can be expressed as $(1+e \, {\rm cos} \, f_{1})^{-2} : (1+e \, {\rm cos} \, f_{2})^{-2}$. So the weight factor $w_{q}$ for each $f_{0,q}$ is given by $w_{q} = (1+e \, {\rm cos} \, f_{0,q})^{-2} / (1+e \, {\rm cos} \, f_{0,19})^{-2}$. $f_{0,19} = 180^{\circ}$ corresponds to the apastron, where the pulsar spends most of its time. Using the values of $\gamma_{m1q}$s, $\gamma_{m2q}$s and $\gamma_{m3q}$s for each $f_{0,q}$, weighted averages can be estimated as:
\begin{equation}
\gamma_{m1,avg} = \frac{\sum_{q} w_{q} \, \gamma_{m1q}}{\sum_{q} w_{q}},
~~~~~~ \gamma_{m2,avg} = \frac{\sum_{q} w_{q} \, \gamma_{m2q}}{\sum_{q} w_{q}},
~~~~~~ \gamma_{m3,avg} = \frac{\sum_{q} w_{q} \, \gamma_{m3q}}{\sum_{q} w_{q}},~~~~~ q = 1, 2 \ldots 36.
\end{equation}
In this paper, we always report these average efficiency factors unless 
otherwise stated. We will henceforth skip the subscript $avg$, in section \ref{sec:res} and subsequent sections. 

\begin{figure}
\centerline{\psfig{figure=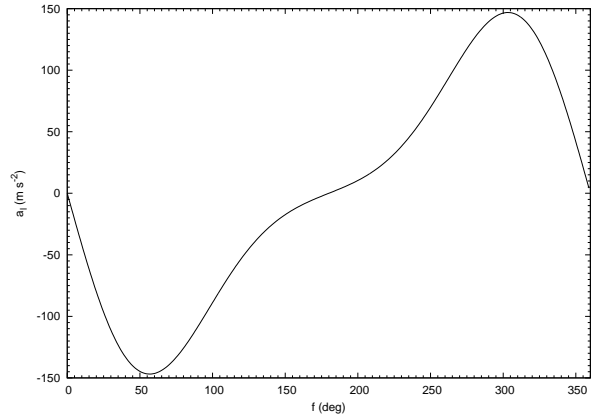,width=8cm,angle=0}}
\caption{Variation of the line-of-sight acceleration with true anomaly over a complete orbit for a binary pulsar with $M_{p}=1.4~M_{\odot}$, $M_{c}=0.3~M_{\odot}$, $i=60^{\circ}$, $\varpi = 0^{\circ}$, $e=0.5$, $P_{o}=0.1$ day. 
} \label{fig:accNSWDwithF}
\end{figure}

In Table \ref{tb:gamma1_dfff}, we show the values of
$\gamma_{m1q}$ for different values of $f_{0}$ and corresponding
values of $w$ for a binary with $M_{p}=1.4~M_{\odot}$,
$M_{c}=0.3~M_{\odot}$, $i=60^{\circ}$, $\varpi = 0^{\circ}$, $e=0.5$,
$P_{o}=0.1$ day, $P_{p} = 0.01$ s, $m=4$, $T = 500$~s. From
Fig. \ref{fig:accNSWDwithF}, we see that $a_l$ has the largest
magnitude at $f=57^{\circ}$.  Both $f_{0}=40^{\circ}$ and
$f_{0}=50^{\circ}$ contains this value of $f$ during the observation
(the pulsar moves $20.83^{\circ}$ during a 500~s long observation), so
we get the smallest value of $\gamma_{m1i}$ here. Note that although
$f_{0}=170^{\circ}$ and $f_{0}=190^{\circ}$ are equally likely (same
value of $w$) and correspond to the same value of $a_l$, but as the
values of $a_l$ during next 500~s are different, they give different
values for $\gamma_{1m}$. Similarly, both $f=0^{\circ}$ and
$f=180^{\circ}$ correspond to $a_l = 0$, but the values of $a_l$
during next 500 sec after $f_{0}=0^{\circ}$ and $f_{0}=180^{\circ}$
are different, so they result different values of $\gamma_{1m}$.

\begin{table*}
\caption{Values of $\gamma_{m1i}$ for different values of $f_{0}$ (true anomaly at the begining of the observation) and corresponding values of $w$ for a binary with $M_{p}=1.4~M_{\odot}$, $M_{c}=0.3~M_{\odot}$, $i=60^{\circ}$, $\varpi = 0^{\circ}$, $e=0.5$, $P_{p} = 0.01$ s, $m=4$, $T = 500$ s. }
\begin{tabular}{ |c|c|c|} \hline
 $f_{0}$ & $\gamma_{1m}$ & $w$ \\ 
\hline
 $0^{\circ}$    & 0.28  & 0.111  \\
 $10^{\circ}$   & 0.23 & 0.112 \\
 $40^{\circ}$   & 0.19 & 0.131  \\
 $50^{\circ}$   & 0.19 & 0.143 \\ 
 $60^{\circ}$   & 0.21 & 0.160 \\
 $70^{\circ}$   & 0.22 & 0.182 \\ 
 $170^{\circ}$  & 0.98 & 0.970 \\  
 $180^{\circ}$  & 0.99 & 1.000 \\ 
 $190^{\circ}$  & 0.86 & 0.970 \\ 
 $350^{\circ}$  & 0.38 & 0.112  \\ \hline
\end{tabular}
\label{tb:gamma1_dfff}
\end{table*}

\section{Results}
\label{sec:res}

\subsection{Neutron Star - White Dwarf (NS-WD) Binaries}
\label{subsec:res_nswd}

If the neutron star is a recycled millisecond pulsar, then the
orbit is expected to be circular because of the strong tidal phase in
the past when the neutron star accreted matter from the
companion \citep{acrs82,radsrini82}. However, in dense stellar environments like globular
clusters or nuclear star clusters near the Galactic center, stellar
interactions lead to eccentric orbits for recycled pulsar
binaries. Many such systems in globular clusters are known at
present\footnote{http://www.naic.edu/$\sim$pfreire/GCpsr.html}. Moreover,
three body interactions can also produce millisecond pulsars in
eccentric orbit. One such system is PSR J1903$+$0327, which was a
member of a triple system in the past, but has become a binary by ejecting one member of the system \citep{crlc08, pvvn11}. 
Fig.~\ref{fig:radialvelnswd} shows the line-of-sight velocity plots for a
canonical example of this category.

\begin{figure*}
 \begin{center}
\hskip -2cm \subfigure[$\varpi = 0^{\circ}$]{\label{subfig:gamma1_omvar1}\includegraphics[width=0.6\textwidth,angle=0]{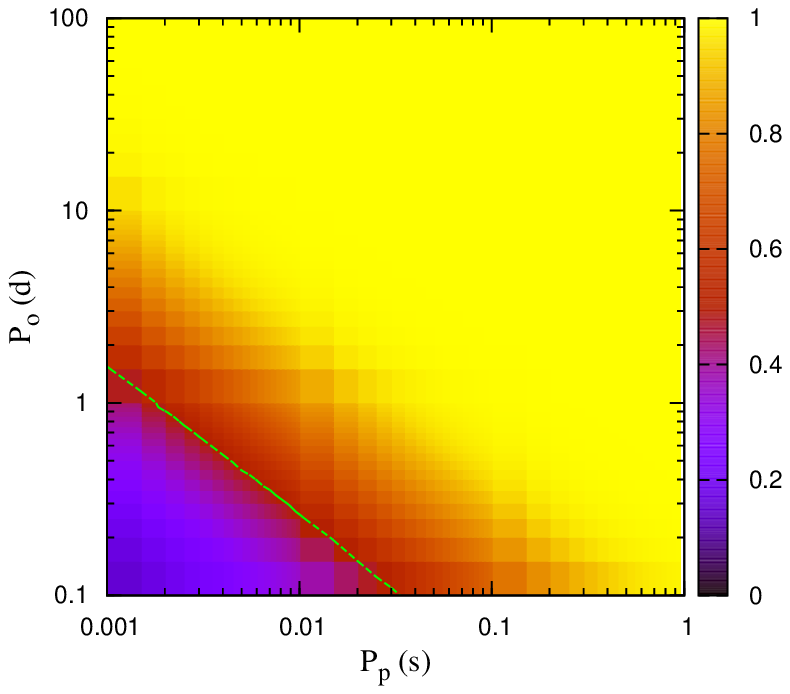}}
\hskip -2cm \subfigure[$\varpi = 30^{\circ}$]{\label{subfig:gamma1_omvar2}\includegraphics[width=0.6\textwidth,angle=0]{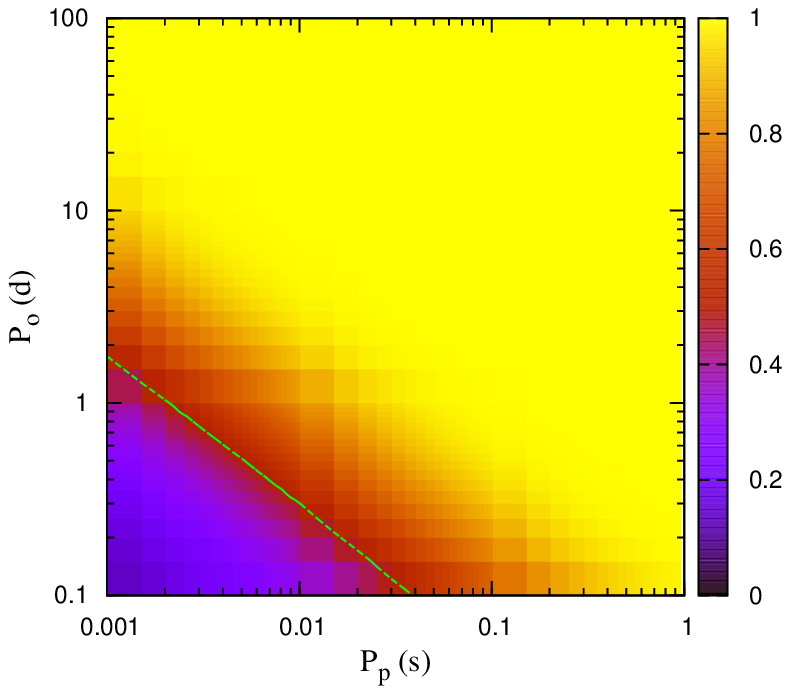}}
\vskip 0.1cm \hskip -2cm \subfigure[$\varpi = 60^{\circ}$]{\label{subfig:gamma1_omvar3}\includegraphics[width=0.6\textwidth,angle=0]{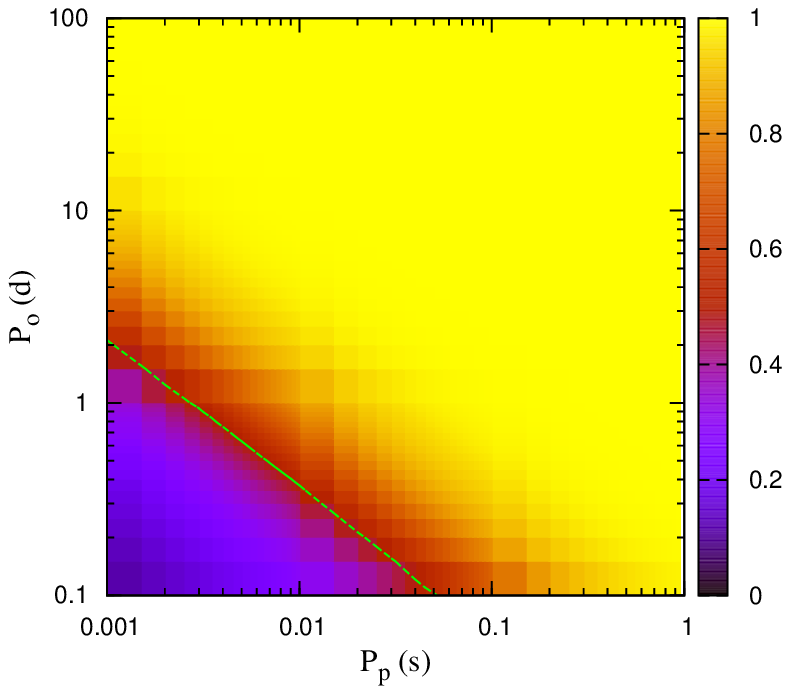}}
\hskip -2cm \subfigure[$\varpi = 90^{\circ}$]{\label{subfig:gamma1_omvar4}\includegraphics[width=0.6\textwidth,angle=0]{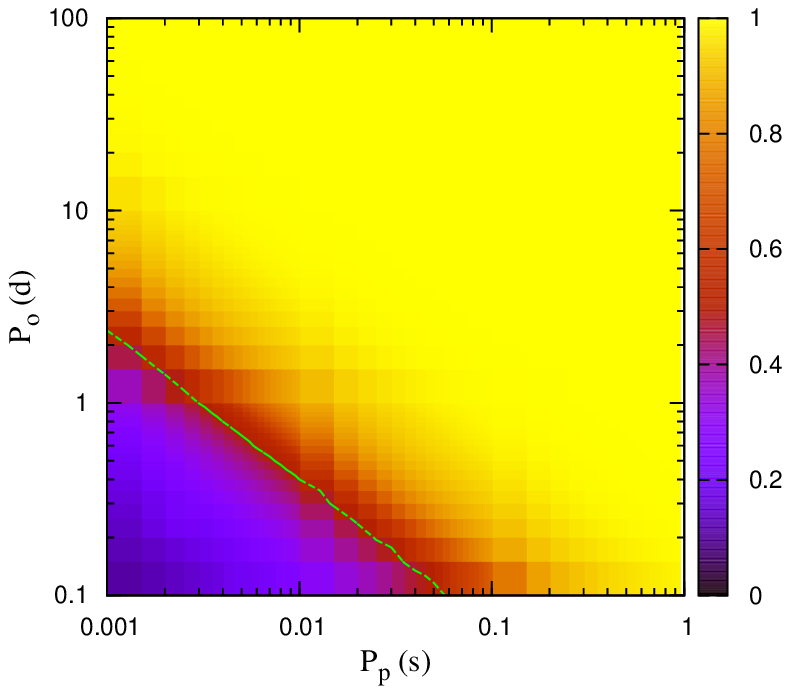}}
 \end{center}
\caption{Variation of $\gamma_{1m}$ with $P_{o}$ and $P_{s}$ for
  different $\varpi$. For each case, $i=60^{\circ}$,
  $M_{p}=1.4~M_{\odot}$, $M_{c}=0.3~M_{\odot}$, $e=0.5$, $m=4$,
  $T=1000$~s.}
\label{fig:gamma1_omvar}
\end{figure*}

\begin{figure*}
 \begin{center}
\hskip -2cm \subfigure[$i = 30^{\circ}$]{\label{subfig:gamma1_ivar1}\includegraphics[width=0.6\textwidth,angle=0]{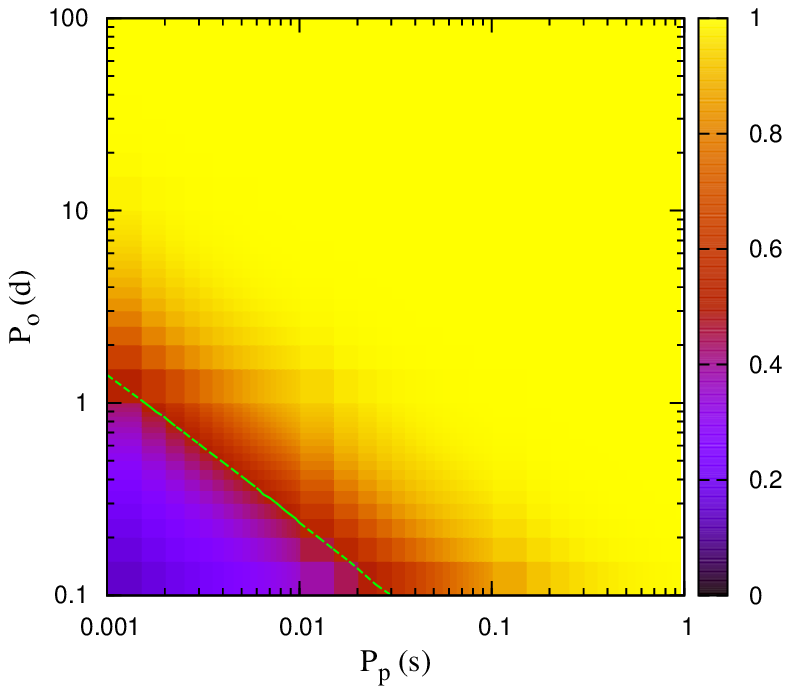}}
\hskip -2cm \subfigure[$i = 90^{\circ}$]{\label{subfig:gamma1_ivar3}\includegraphics[width=0.6\textwidth,angle=0]{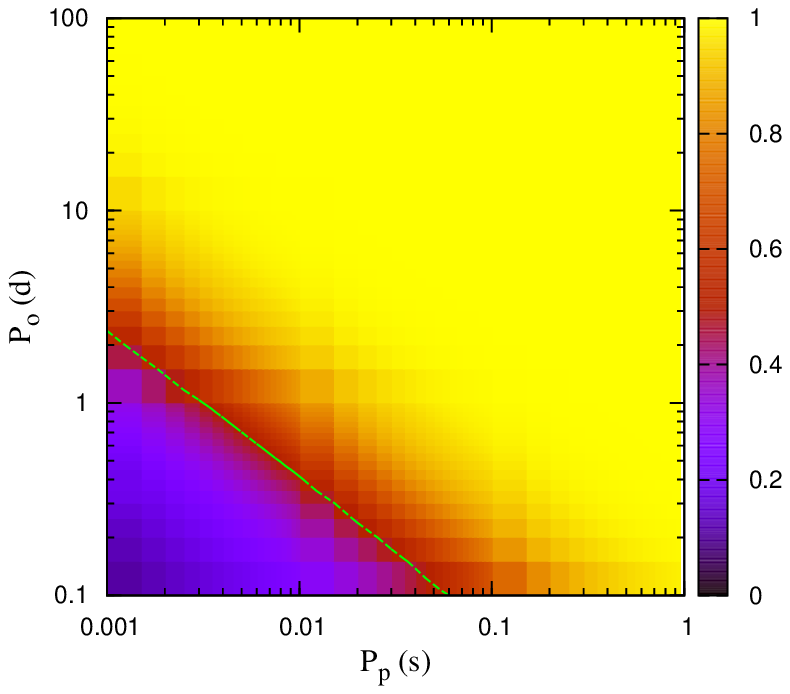}}
 \end{center}
\caption{Variation of $\gamma_{1m}$ with $P_{o}$ and $P_{s}$ for
  different $i$. For each case, $\varpi=60^{\circ}$,
  $M_{p}=1.4~M_{\odot}$, $M_{c}=0.3~M_{\odot}$, $e=0.5$, $m=4$,
  $T=1000$~s.}
\label{fig:gamma1_ivar}
\end{figure*}

\begin{figure*}
 \begin{center}
\hskip -2cm \subfigure[$M_{c}=0.01~M_{\odot}$]{\label{subfig:gamma1_Mcvar2}\includegraphics[width=0.6\textwidth,angle=0]{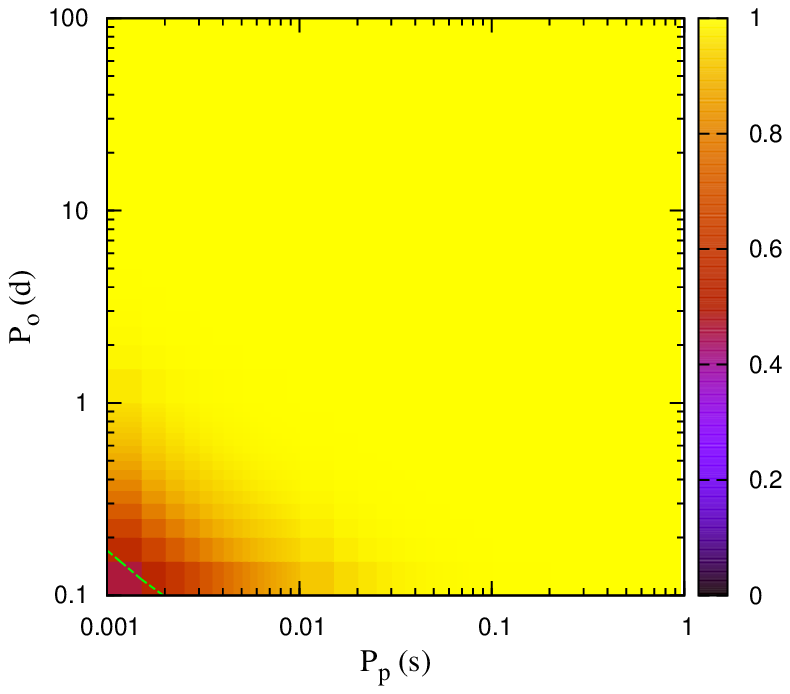}}
\hskip -2cm \subfigure[$M_{c}=0.1~M_{\odot}$]{\label{subfig:gamma1_Mcvar3}\includegraphics[width=0.6\textwidth,angle=0]{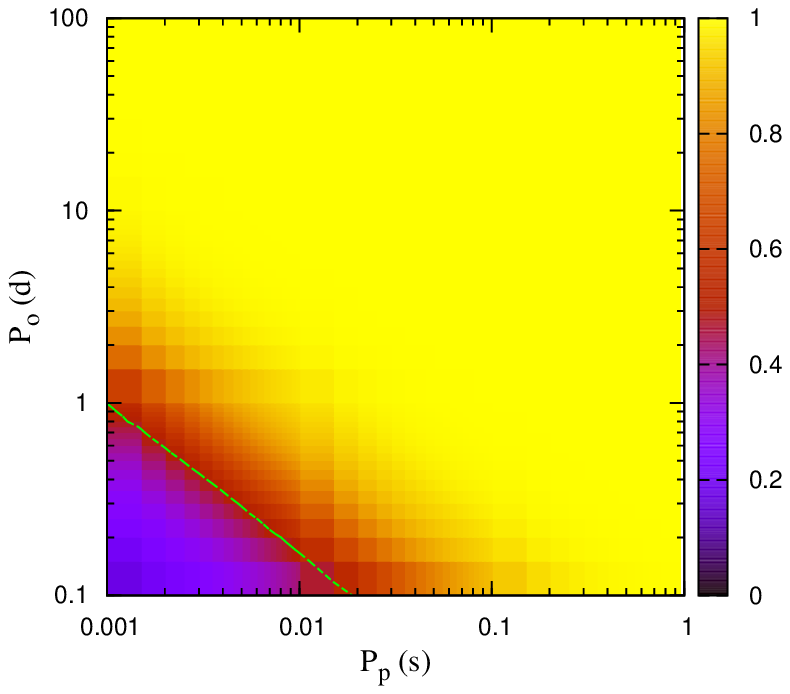}}
\hskip -2cm \subfigure[$M_{c}=0.8~M_{\odot}$]{\label{subfig:gamma1_Mcvar4}\includegraphics[width=0.6\textwidth,angle=0]{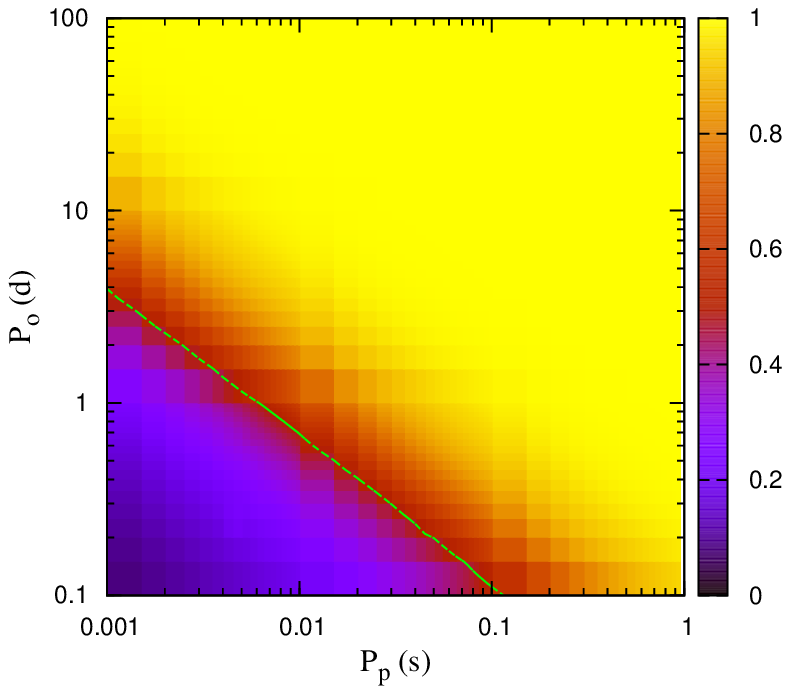}}
 \end{center}
\caption{Variation of $\gamma_{1m}$ with $P_{o}$ and $P_{s}$ for
  different $M_{c}$. For each case, $i=60^{\circ}$,
  $\varpi=60^{\circ}$, $M_{p}=1.4~M_{\odot}$, $e=0.5$, $m=4$,
  $T=1000$~s.}
\label{fig:gamma1_Mcvar}
\end{figure*}

\begin{figure*}
 \begin{center}
\hskip -2cm \subfigure[$M_{p}=~1.0M_{\odot}$]{\label{subfig:gamma1_Mpvar1}\includegraphics[width=0.6\textwidth,angle=0]{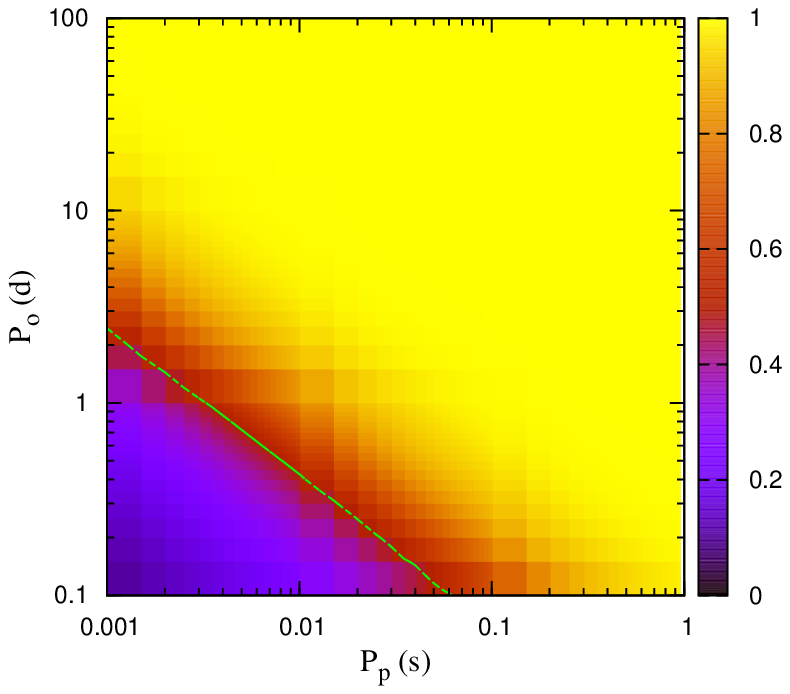}}
\hskip -2cm \subfigure[$M_{p}=2.0~M_{\odot}$]{\label{subfig:gamma1_Mpvar5}\includegraphics[width=0.6\textwidth,angle=0]{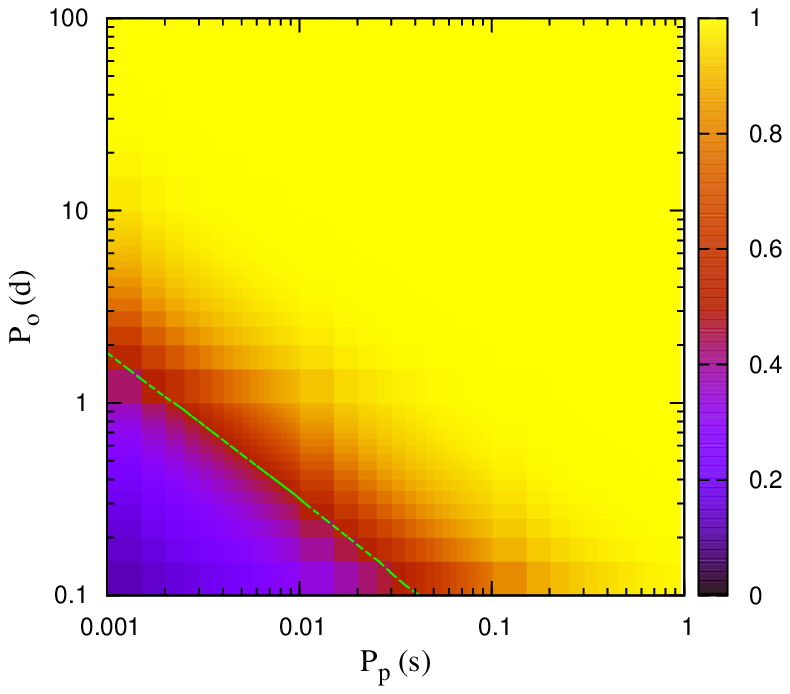}}
 \end{center}
\caption{Variation of $\gamma_{1m}$ with $P_{o}$ and $P_{s}$ for
  different $M_{p}$. For each case, $i=60^{\circ}$,
  $\varpi=60^{\circ}$, $M_{c}=0.3~M_{\odot}$, $e=0.5$, $m=4$,
  $T=1000$~s.}
\label{fig:gamma1_Mpvar}
\end{figure*}

\begin{figure*}
 \begin{center}
\hskip -2cm \subfigure[$e = 0.001$]{\label{subfig:gamma1_eccvar1}\includegraphics[width=0.6\textwidth,angle=0]{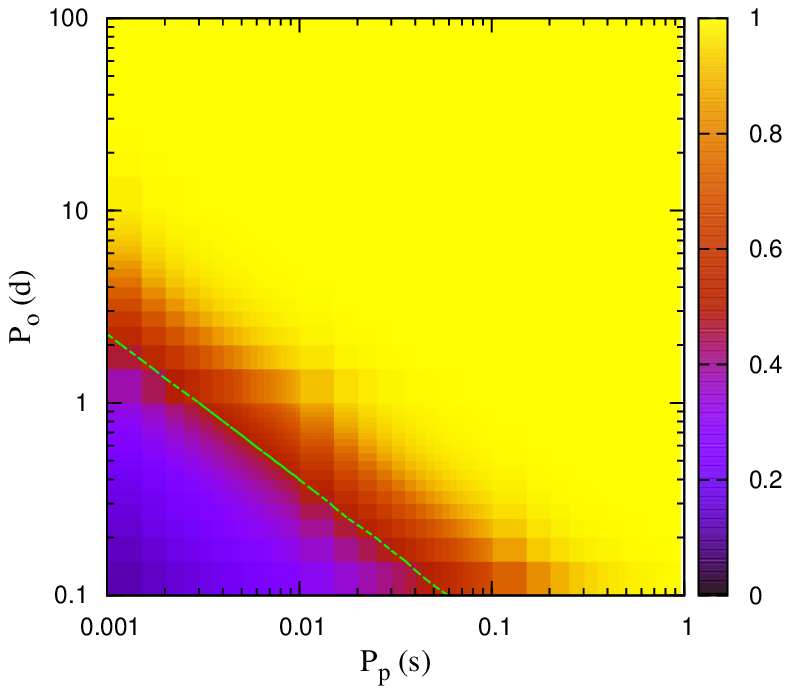}}
\hskip -2cm \subfigure[$e = 0.01$]{\label{subfig:gamma1_eccvar2}\includegraphics[width=0.6\textwidth,angle=0]{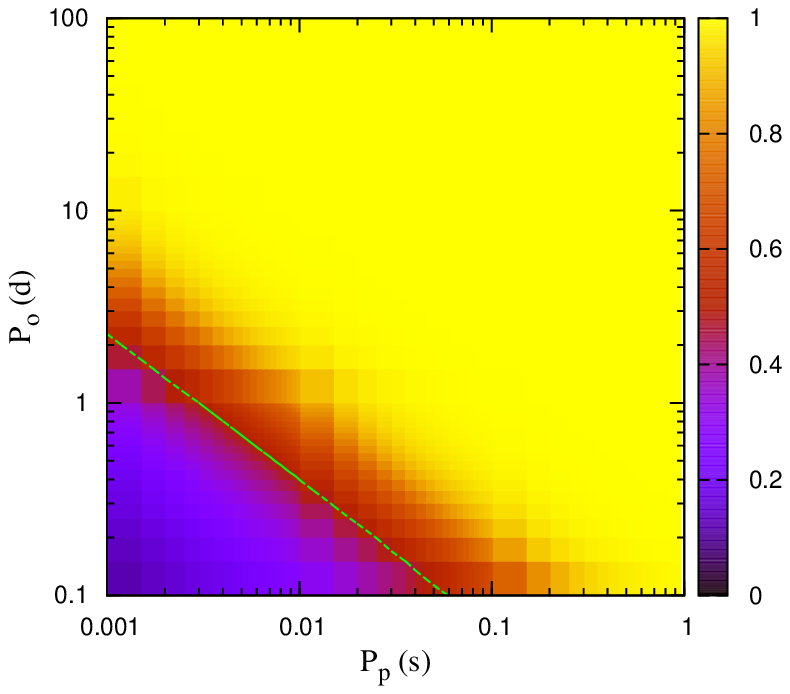}}
\vskip 0.1cm \hskip -2cm \subfigure[$e = 0.1$]{\label{subfig:gamma1_eccvar3}\includegraphics[width=0.6\textwidth,angle=0]{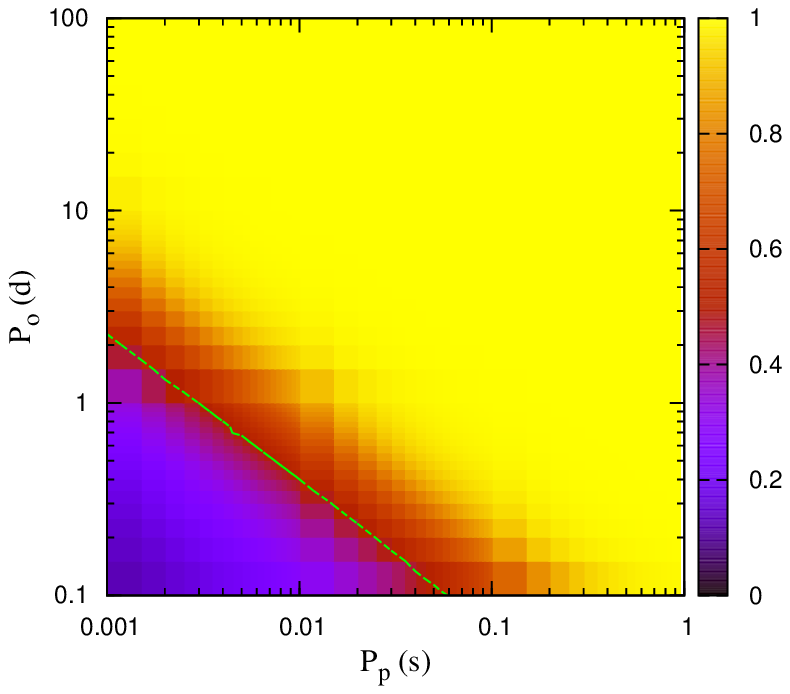}}
\hskip -2cm \subfigure[$e = 0.3$]{\label{subfig:gamma1_eccvar4}\includegraphics[width=0.6\textwidth,angle=0]{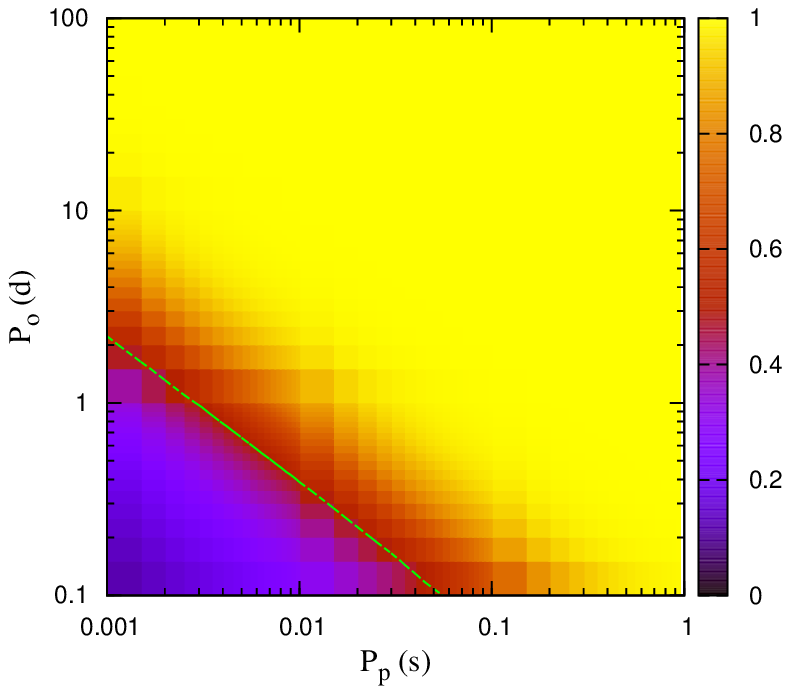}}
\vskip 0.1cm \hskip -2cm \subfigure[$e=0.5$]{\label{subfig:gamma1_eccvar5}\includegraphics[width=0.6\textwidth,angle=0]{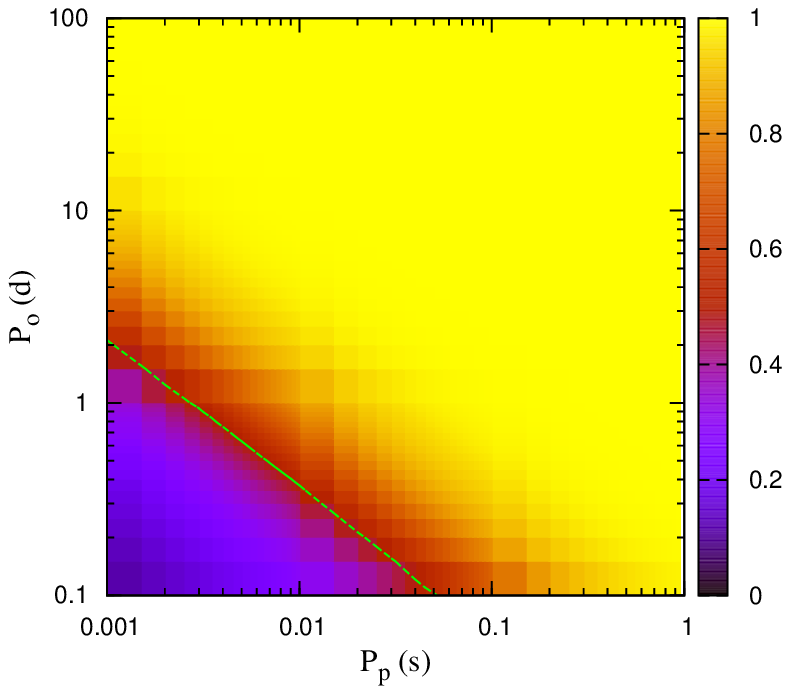}}
\hskip -2cm \subfigure[$e=0.8$]{\label{subfig:gamma1_eccvar8}\includegraphics[width=0.6\textwidth,angle=0]{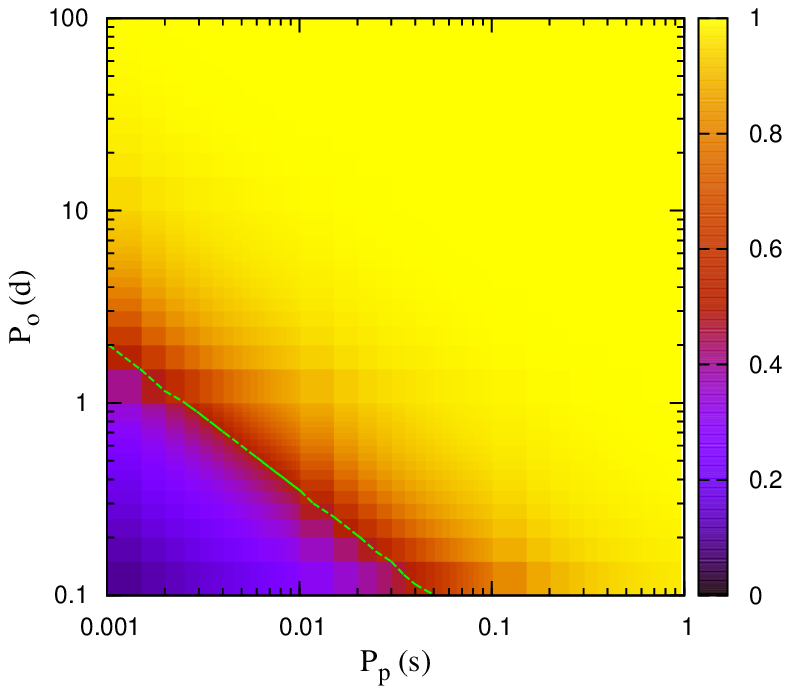}}
 \end{center}
\caption{Variation of $\gamma_{1m}$ with $P_{o}$ and $P_{s}$ for different $e$. For each case, $i=60^{\circ}$, $\varpi=60^{\circ}$, $M_{p}=1.4~M_{\odot}$, $M_{c}=0.3~M_{\odot}$, $m=4$and $T=1000$~s. }
\label{fig:gamma1_eccvar}
\end{figure*}

\begin{figure*}
 \begin{center}
\hskip -2cm \subfigure[$m = 1$]{\label{subfig:gamma1_mvar1}\includegraphics[width=0.6\textwidth,angle=0]{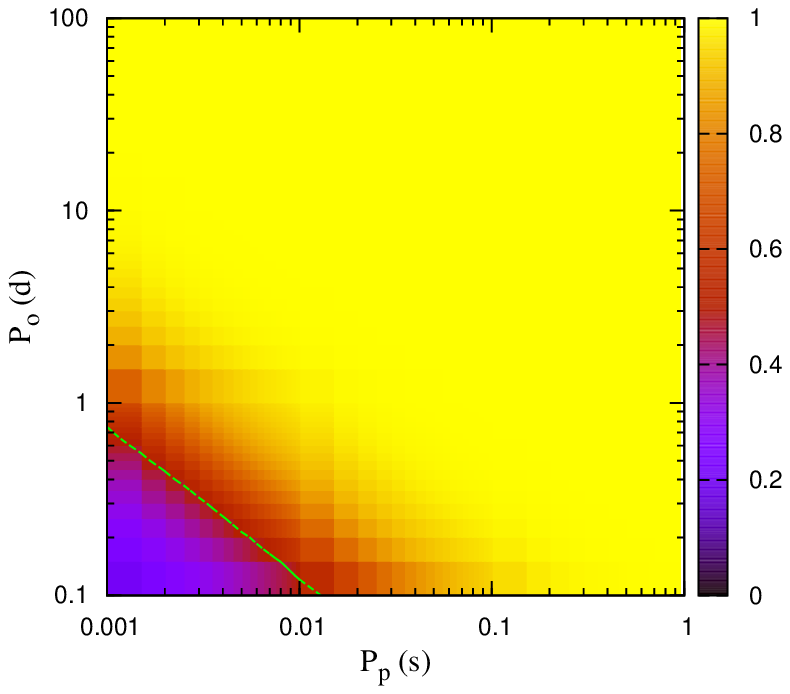}}
\hskip -2cm \subfigure[$m = 7$]{\label{subfig:gamma1_mvar2}\includegraphics[width=0.6\textwidth,angle=0]{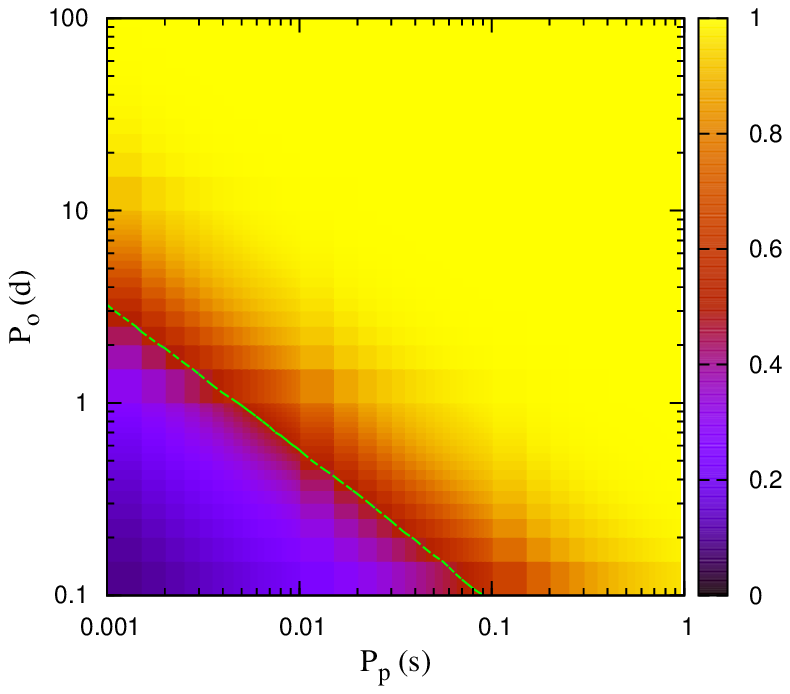}}
 \end{center}
\caption{Variation of $\gamma_{1m}$ with $P_{o}$ and $P_{s}$ for different $m$. For each case, $\varpi=60^{\circ}$, $i=60^{\circ}$, $M_{p}=1.4~M_{\odot}$, $M_{c}=0.3~M_{\odot}$, $e=0.5$and $T=1000$~s.}
\label{fig:gamma1_mvar}
\end{figure*}

\begin{figure*}
 \begin{center}
\hskip -2cm \subfigure[$T = 500$ s]{\label{subfig:gamma1_Tvar1}\includegraphics[width=0.6\textwidth,angle=0]{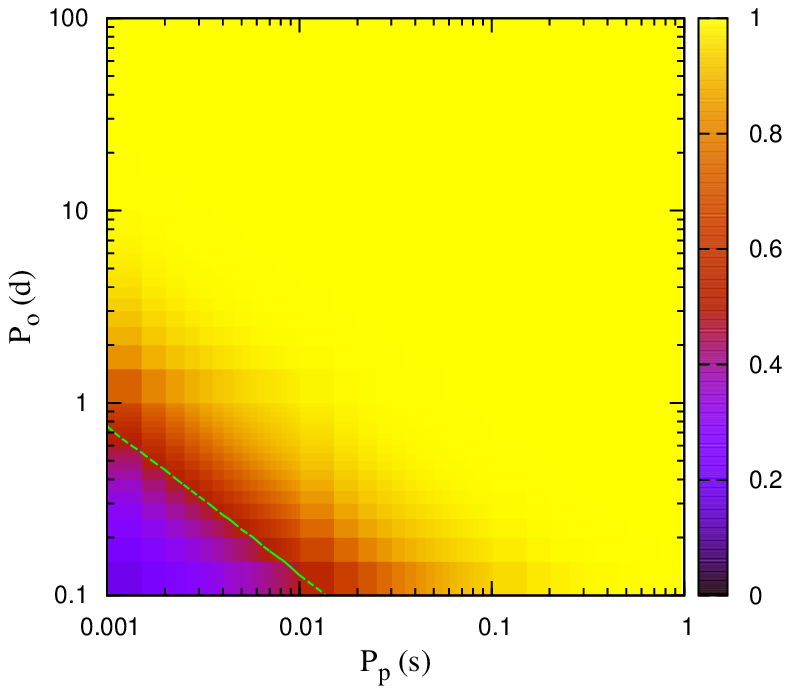}}
\hskip -2cm \subfigure[$T = 2000$ s]{\label{subfig:gamma1_Tvar2}\includegraphics[width=0.6\textwidth,angle=0]{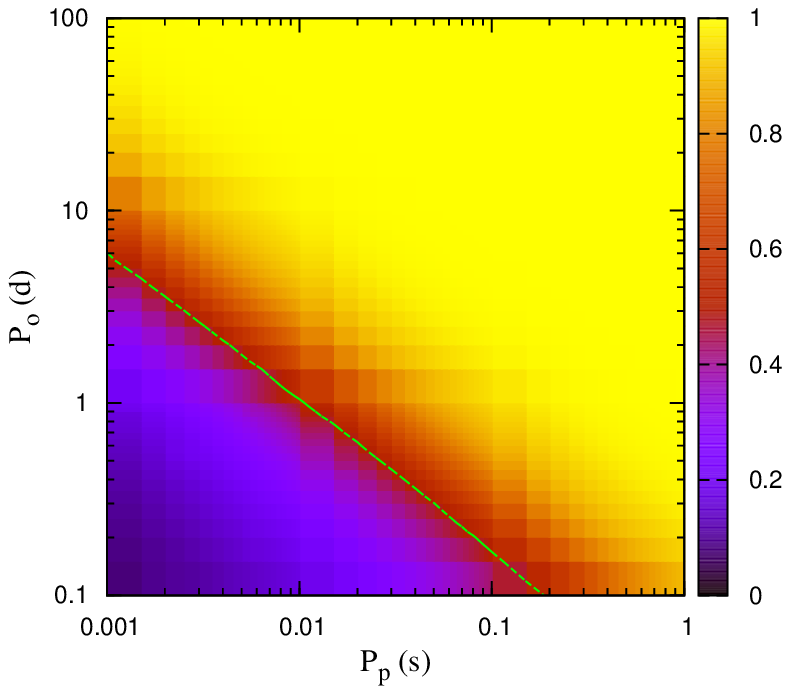}}
 \end{center}
\caption{Variation of $\gamma_{1m}$ with $P_{o}$ and $P_{s}$ for different $T$. For each case, $\varpi=60^{\circ}$, $i=60^{\circ}$, $M_{p}=1.4~M_{\odot}$, $M_{c}=0.3~M_{\odot}$, $e=0.5$, $m = 4$.  }
\label{fig:gamma1_Tvar}
\end{figure*}

It is clear from Equation (\ref{eq:ob_angfreq}), that the spread of
the power among adjacent Fourier bins is larger for higher values of
$m$ or larger values of $\omega_{p}$, thus the values of the
efficiency factors decrease for higher $m$ or smaller
$P_{p}$. Similarly, a smaller value of $P_{o}$ gives higher values of
$v_{l}$, $a_{l}$ and $j_{l}$ and so on (all other parameters being
fixed) resulting in a larger spread of the power among adjacent Fourier
bins, \textit{i.e.}, reduction of the efficiency factors. It is
evident from Eqn. (\ref{eq:keplerapPo}) that higher value of $\sin i$
will increase the values of $v_{l}$, $a_{l}$, $j_{l}$ and so on, leading
to a decrease in the efficiency factors. To demonstrate these facts
graphically, we show variations of $\gamma_{1m}$ with with $P_{o}$ and
$P_{s}$ for different parameters. In all the subsequent plots, the X-axis
shows the values of $P_{p}$, the Y-axis shows the values of $P_{o}$
and the color code represents the values of $\gamma_{1m}$. The contour
for $\gamma_{1m}=0.5$ is also shown in each plot. We can say that the
pulsars having $\gamma_{1m} < 0.5$ (lying on the bottom-left side of the
$\gamma_{1m}=0.5$ contour) are difficult to detect. Fig. \ref{fig:gamma1_omvar} shows the variation for different
values of $\varpi$. We see that the $\gamma_{1m}=0.5$ contour shifts
rightward with the increase of $\varpi$, making a larger portion of the
phase-space difficult to detect. Fig. \ref{fig:gamma1_ivar} shows the
variation for different values of $i$, and as expected, the increase
of $i$ makes a larger portion of the phase-space difficult to
detect. In Fig. \ref{fig:gamma1_Mcvar}, we show variations of
$\gamma_{1m}$ with with $P_{o}$ and $P_{s}$ for different $M_c$
keeping all other parameters fixed, and we see that the values of
$\gamma_{1m}$ are lower for higher values of $M_c$ when all other
parameters are the same. This happens because $a_p$ increases with the
increase of $M_c$ (if $P_o$ remains the same) giving higher values of
$v_{l}$, $a_{l}$ and $j_{l}$ etc. resulting in a larger spread of the power
among adjacent Fourier bins, \textit{i.e.}, reduction of the
efficiency factors. As an example, if $M_c$ increases from
$0.1~{M_{\odot}}$ to $0.8~{M_{\odot}}$, the fractional increase in $a_p$
is 5.2.  On the other hand, from Fig \ref{fig:gamma1_Mpvar}, we have
not seen any significant effect of the variation of $M_p$ in the
realistic range of $1 - 2 ~{M_{\odot}}$, because the fractional
increase of $a_p$ is only 0.32 when $M_p$ increases from
$1~{M_{\odot}}$ to $2 ~{M_{\odot}}$. In Fig. \ref{fig:gamma1_eccvar},
we show variations of $\gamma_{1m}$ with with $P_{o}$ and $P_{s}$ for
different values of $e$ and find that it does not play any significant
role unless very high when $\gamma_{1m}$ increases with the increase
of $e$. In Fig. \ref{fig:gamma1_mvar}, we show the variations of
$\gamma_{1m}$ with with $P_{o}$ and $P_{s}$ for $m=1$ and $m=7$
keeping all the other parameters fixed and, as expected, $\gamma_{1m}$
decreases for higher values of $m$. In Fig. \ref{fig:gamma1_Tvar}, we
show variations of $\gamma_{1m}$ with with $P_{o}$ and $P_{s}$ for
different values of $T$ and find that a smaller value of $T$ makes a
larger portion of the phase space easy to detect. This fact was also noticed by \citet{rzw98}. On the other hand,
smaller $T$ reduces the value of minimum detectable flux density for a
pulsar, independent of its orbital parameters \citep{lk05}. Hence an
optimal strategy is needed to choose a value of $T$ depending upon the
motivation of the survey and efficiency of the telescope.

%%%%%%%%%%%%%%%%%%%%%%%%%%%%%%%%%%%%%%%%%%%%

%%%%%%%%%%%%%%%%%%%%%%%%%%%%%%%%%%%%%%%%%%%%
%%%%%%%%%%%%%%%%%%%%%%%%%%%%%%%%%%%%%%%%%%%%
%%%%%%%%%%%%%%%%%%%%%%%%%%%%%%%%%%%%%%%%%%%%
%%%%%%%%%%%%%%%%%%%%%%%%%%%%%%%%%%%%%%%%%%%%
%%%%%%%%%%%%%%%%%%%%%%%%%%%%%%%%%%%%%%%%%%%%
%%%%%%%%%%%%%%%%%%%%%%%%%%%%%%%%%%%%%%%%%%%%
%%%%%%%%%%%%%%%%%%%%%%%%%%%%%%%%%%%%%%%%%%%%
%%%%%%%%%%%%%%%%%%%%%%%%%%%%%%%%%%%%%%%%%%%%

%%%%%%%%%%%%%%%%%%%%%%%%%%%%%%%%%%%%%%%%%%%%
%%%%%%%%%%%%%%%%%%%%%%%%%%%%%%%%%%%%%%%%%%%%

%%%%%%%%%%%%%%%%%%%%%%%%%%%%%%%%%%%%%%%%%%%%
%%%%%%%%%%%%%%%%%%%%%%%%%%%%%%%%%%%%%%%%%%%%

\begin{figure*}
 \begin{center}
\hskip -2cm \subfigure[$\gamma_{2m}$]{\label{subfig:gamma2_NSWD}\includegraphics[width=0.6\textwidth,angle=0]{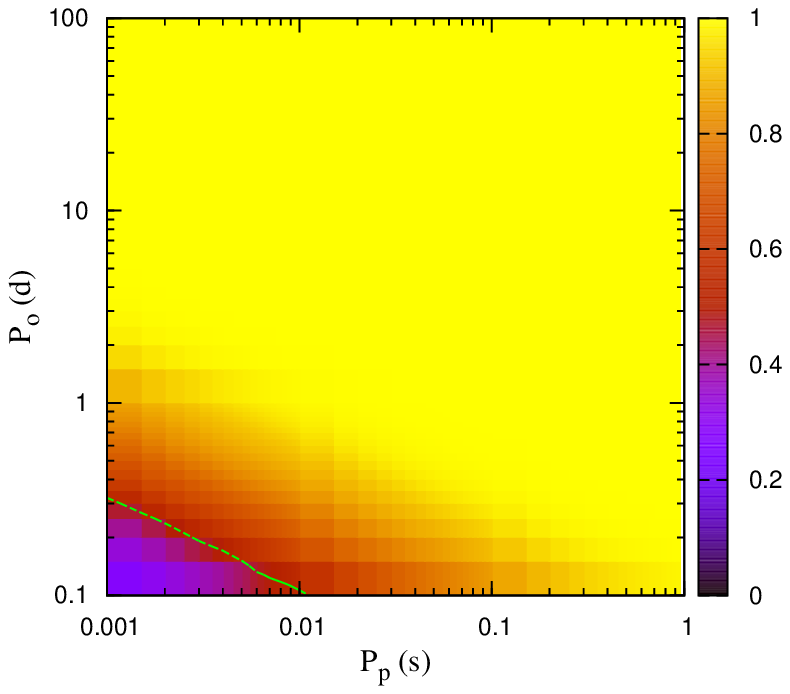}}
\hskip -2cm \subfigure[$\gamma_{3m}$]{\label{subfig:gamma3_NSWD}\includegraphics[width=0.6\textwidth,angle=0]{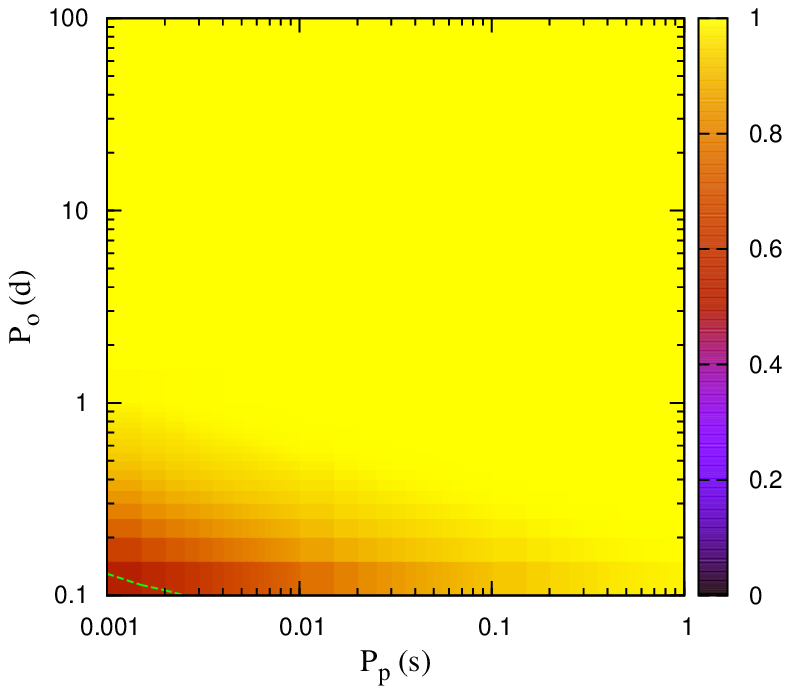}}
 \end{center}
\caption{Variations of $\gamma_{2m}$ and $\gamma_{3m}$ with $P_{o}$ and $P_{s}$ keeping $\varpi=30^{\circ}$, $i=60^{\circ}$ $M_{p}=1.4~M_{\odot}$, $M_{c}=0.3~M_{\odot}$, $m=4$ and $T=1000$ s. }
\label{fig:gamma2o3}
\end{figure*}
%%%%%%%%%%%%%%%%%%%%%%%%%%%%%%%%%%%%%%%%%%%%

In Fig. \ref{fig:gamma2o3}, we show variation of $\gamma_{2m}$ and
$\gamma_{3m}$ with $P_{o}$ and $P_{s}$ keeping $\varpi=30^{\circ}$,
$i=60^{\circ}$ $M_{p}=1.4~M_{\odot}$, $M_{c}=0.3~M_{\odot}$, $m=4$ and
$T=1000$ s. These can be compared to Fig. \ref{subfig:gamma1_omvar2}
which shows the variation of $\gamma_{1m}$ for the same set of
parameters. The increases in efficiency factors by using
acceleration or acceleration-jerk searches are significant. We
notice that the slope of the contour is different and the use of
acceleration or acceleration-jerk searches are more effective
for short orbital periods. This result agrees with that obtained by JK91.

In Table \ref{tb:nswd_gammas}, we present the efficiency factors for
$T=1000$~s and $m=4$, averaged over different $f_0$s as described
earlier for the NS-WD binaries for which all the required parameters,
e.g., $P_{p}$, $e$, $P_{o}$, $\sin i$, $\varpi$, $M_c$ and $M_p$ are
known, for few cases, the values of $M_p$ are not known, so assumed to
be $1.4~M_{\odot}$. Table \ref{tb:nswd_params} shows the parameters
for these binaries and corresponding references.

\begin{table*}
\caption{Parameters for NS-WD binaries for which we can calculate efficiency factors. The columns from left to right show the name of the pulsar, spin period, orbital period, orbital eccentricity, sine of the inclination angle, longitude of periastron, companion mass, pulsar mass and corresponding references. }
\begin{tabular}{|l|l|l|l|l|l|l|l|l|} \hline \hline 
Pulsar & $P_{p}$  & $P_{o}$  & $e$ & $\sin i$ & $\varpi$ & $M_{c}$ & $M_{p}$ & Refs.\\  
 & (sec)  & (day)  &  &  & (deg) & ($M_{\odot}$) & ($M_{\odot}$) & \\  \hline 
J0437$-$4715 & 0.005757  & 5.74105  &  1.918 $\times 10^{-5}$ & 0.674  &  1.222  & 0.254  & 1.76  & \citet{vbv08}\\  
J0751$+$1807 & 0.003479   &  0.26314   & 5.0 $\times 10^{-7}$ &  0.9121 & 45.0   &  0.191 & 1.26  & \citet{nss05};  ATNF catlogue$^\ddagger$ for $\varpi$ \\ 
 &    &    &  &   &    &   &   &  \citet[for $M_{p}$]{nsk07} \\ 
J1600$-$3053 & 0.003598 & 14.34846  & 17.369 $\times 10^{-5}$ &  0.8 & 181.85  & 0.6 & 1.4$^*$  & \citet{vbc09} \\ 
J1603$-$7202 & 0.014842  &  6.30863 & 9.3 $\times 10^{-6}$ & 0.89  & 169  & 0.14 & 1.4$^*$  & \citet{hbo06} \\ 
J1614$-$2230 & 0.003151  &  8.68662  & 1.30 $\times 10^{-6}$ &  0.9999 & 175.1  & 0.500  &  1.97  & \citet{dpr10}\\ 
J1640$+$2224 & 0.003163   & 175.46066  & 7.973 $\times 10^{-4}$ &  0.99 & 50.731  & 0.15 & 1.4$^*$  & \citet{llww06} \\ 
J1713$+$0747 & 0.00457  & 67.82513  & 7.494 $\times 10^{-5}$ & 0.95  & 176.191  & 0.28 & 1.3 & \citet{snslb05} \\ 
J1738$+$0333 & 0.00585   & 0.35479  & 3.4 $\times 10^{-7}$ & 0.5388   & 155.695 & 0.19 & 1.46 & \citet{pwf12} \\
J1802$-$2124 & 0.012648   & 0.69889  & 2.48 $\times 10^{-6}$ & 0.9845   & 20.0 & 0.78 & 1.24 & \citet{fsk10} \\
J1903$+$0327 &  0.00215  & 95.17412  & 0.4367 &  0.976 &  141.652 & 1.03 &  1.67 & \citet{fbw11}\\  \hline \hline
\end{tabular}
\label{tb:nswd_params}
\vskip 0.1 cm
{\footnotesize{$^\ddagger$ : http://www.atnf.csiro.au/research/pulsar/psrcat/expert.html \\} 
\footnotesize{$^*$ : $M_{p}$ not measured, taken as $1.4~{\rm M_{\odot}}$}
}
\end{table*}

\begin{table*}
\caption{Efficiency factors and corresponding parameters for NS-WD binaries (for which all required parameters are known) for $T=1000$ s, $m=4$ (averaged over different $T_{p}$s as described in the text). }
\begin{tabular}{|l|c|c|c|} \hline \hline 
pulsar & $\gamma_{1m}$  & $\gamma_{2m}$  &  $\gamma_{3m}$   \\  \hline 
J0437$-$4715 &  1.00  &  1.00   &  1.00   \\  
J0751$+$1807 &  0.29  &  0.64   &  0.98   \\
J1600$-$3053 &  1.00  &  1.00   &  1.00   \\ 
J1603$-$7202 &  1.00  &  1.00   &  1.00   \\ 
J1614$-$2230 &  0.99  &  1.00   &  1.00   \\ 
J1640$+$2224 &  1.00  &  1.00   &  1.00   \\ 
J1713$+$0747 &  1.00  &  1.00   &  1.00   \\ 
J1738$+$0333 &  0.56  &  0.95   &  1.00   \\ 
J1802$-$2124 &  0.50  &  0.98   &  1.00    \\
J1903$+$0327 &  1.00  &  1.00   &  1.00   \\  \hline \hline
\end{tabular}
\label{tb:nswd_gammas}
\end{table*}

For the sake of simplicity, in most cases, we compute the efficiency factors for the $m=4$ harmonic, but the procedure will be the same for any other harmonic. The value of the efficiency factor at any harmonic $m_{2}$
for spin period $P_{p_{2}}$ would be the same as that at the harmonic $m_{1}$ and spin period $P_{p_{1}}$ if $P_{p_{2}} = \frac{m_{2}}{m_{1}} P_{p_{1}}$ if we keep all other parameters unchanged. This fact can be used to read the efficiency factors for any other harmonic from the plots.

In the case of a real pulsar, the number of harmonics present in the Fourier spectrum of a pulse is roughly reciprocal of the pulse duty cycle \citep{lk05}. Harmonic summing is used to get the power from each of these harmonics. The power of lower harmonics are higher as well as the efficiency factors. As a combined effect, the contributions from higher harmonics become more significant for a binary pulsar. An an example, we fit the pulse profile of PSR J1802$-$2124 \citep{fsk04} with a Gaussian, and compute the Fourier spectrum. Table \ref{tb:nswd_harmonicsumming} shows the values of $\gamma_{1m}$ for first 10 harmonics, power ($D_{m}$) of each harmonic (in arbitrary unit) and degraded power ($\gamma_{1m}^2 \, D_{m}$) of each harmonic for different observation duration, e.g. $T=$500, 100, 1500 and 2000 s. As expected, for any fixed value of $T$, $\gamma_{1m}$ decreases with increasing value of $m$, and for any fixed value of $m$,  $\gamma_{1m}$ decreases with the increase of $T$. Note that, for any fixed value of $T$, the decrease in the value of $\gamma_{1m}^2 \, D_{m}$ from $m=1$ to $m=10$ is slower than that of $D_{m}$.

\begin{table*}
\caption{Values of $\gamma_{1m}$ for first 10 harmonics of PSR J1802$-$2124, power ($D_{m}$) of each harmonic (in arbitrary unit) and degraded power ($\gamma_{1m}^2 \, D_{m}$) of each harmonic for $T=$500, 100, 1500 and 2000 s. The pulse profile was fitted with a Gaussian.}
\begin{tabular}{|c|c|c|c|c|c|c|c|c|c|} \hline \hline 
    &  & \multicolumn{2}{|c|}{$T=500$ s} & \multicolumn{2}{|c|}{$T=1000$ s} & \multicolumn{2}{|c|}{$T=1500$ s} & \multicolumn{2}{|c|}{$T=2000$ s}\\ 
harmonic number ($m$)    & power ($D_{m}$) & $\gamma_{1m}$ & $\gamma_{1m}^2 \, D_{m}$ & $\gamma_{1m}$ & $\gamma_{1m}^2 \, D_{m}$ & $\gamma_{1m}$ & $\gamma_{1m}^2 \, D_{m}$ & $\gamma_{1m}$ & $\gamma_{1m}^2 \, D_{m}$\\  \hline 
1        &   1.00000 & 0.99  & 0.982998  & 0.87 & 0.761651 & 0.64  & 0.412938  & 0.51  & 0.257970  \\
2        &   0.88965 & 0.97  & 0.830733  & 0.67 & 0.400844 & 0.48  & 0.206042  & 0.38  & 0.128296 \\ 
3        &   0.62585 & 0.93  & 0.536621  & 0.57 & 0.203214 & 0.41  & 0.103509  & 0.31  & 0.060903  \\
4        &   0.34728 & 0.87  & 0.264492  & 0.50 & 0.088176 & 0.36  & 0.043970  & 0.27  & 0.025762 \\
5        &   0.15156 & 0.81  & 0.099354  & 0.46 & 0.031453 & 0.32  & 0.015686  & 0.25  & 0.009186  \\
6        &   0.05199 & 0.74  & 0.028554  & 0.42 & 0.009198 & 0.30  & 0.004363  & 0.23  & 0.002693  \\
7        &   0.01412 & 0.70  & 0.006924  & 0.39 & 0.002202 & 0.27  & 0.001009  & 0.22  & 0.000674  \\
8        &   0.00310 & 0.67  & 0.001394  & 0.38 & 0.000441 & 0.25  & 0.000196  & 0.20  & 0.000125 \\
9        &   0.00056 & 0.64  & 0.000230  & 0.35 & 0.000070 & 0.24  & 0.000032  & 0.19  & 0.000020 \\
10       &   0.00008 & 0.61  & 0.000029  & 0.34 & 0.000009 & 0.23  & 0.000004  & 0.18  & 0.000003  \\  \hline \hline
\end{tabular}
\label{tb:nswd_harmonicsumming}
\end{table*}

\subsection{Double Neutron Star (DNS) Binaries}
\label{subsec:res_nsns}

Presently there are eight confirmed double neutron star (DNS) binaries and four more candidate DNSs. Among these twelve, all required parameters to calculate efficiency factors are known only for four. We present the efficiency factors for these DNSs in Table \ref{tb:dns_gammas}, for $T=1000$ s, $m=4$. The parameters for these DNSs with corresponding references are given in Table \ref{tb:dns_params}.

\begin{table*}
\caption{Parameters for the DNSs for which we can calculate efficiency factors. The columns from left to right show the name of the pulsar, spin period, orbital period, orbital eccentricity, sine of the inclination angle, longitude of periastron, companion mass, pulsar mass and corresponding references.}
\begin{tabular}{lllllllll} \hline \hline 
DNS & $P_{p}$  & $P_{o}$  & $e$ & $\sin i$ & $\varpi$ & $M_{c}$ & $M_{p}$ & Refs.\\  
 & (sec)  & (day)  &  &  & (deg) & ($M_{\odot}$) & ($M_{\odot}$) & \\  \hline 
J0737$-$3039A & 0.022699  & 0.10225   & 0.08778   & 0.9997 & 87.033 & 1.249(B) & 1.338(A) & \citet{ksm06} \\ 
J0737$-$3039B & 2.773461  & 0.10225   & 0.08778   & 0.9997 & 267.033 & 1.338(A) &  1.249(B) & \citet{ksm06} \\
B1534+12 & 0.037904   & 0.42074  &  0.27368  & 0.975 & 274.769  & 1.35  & 1.328 & \citet{kws03}; \citet[for $M_{tot}$]{stairs02} \\ 
J1756$-$2251 & 0.028461 & 0.31963  & 0.18057 & 0.95 & 327.825 & 1.312 & 1.258 & \citet{ferd08}\\ 
J1807$-$2500B$\dagger$ & 0.004186  & 9.95667  &  0.74703  & 0.996   & 11.335 & 1.206 &  1.365 & \citet{lfrj12} \\  
B1913+16 & 0.059030   & 0.32300  & 0.61713 & 0.71  &  292.545  &  1.389 & 1.440 & \citet{wnt10};  \\ 
 &    &   &  &   &    &   &  &   \citet[for $\sin i$]{ds88} \\  \hline \hline
\end{tabular}
\label{tb:dns_params}
\vskip 0.1 cm
{\footnotesize{$^\dagger$ : candidate DNS. } }
\end{table*}

\begin{table*}
\caption{Efficiency factors for DNSs (for which all required parameters are known) for $T=1000$ s, $m=4$. }
\begin{tabular}{lccc} \hline \hline 
DNS & $\gamma_{1m}$  & $\gamma_{2m}$  &  $\gamma_{3m}$   \\  \hline 
J0737$-$3039A & 0.21  &  0.41  &  0.56 \\ 
J0737$-$3039B & 0.98  &  1.00  &  1.00 \\
B1534+12      & 0.52  &  0.92  &  1.00 \\ 
J1756$-$2251  & 0.41  &  0.85  &  0.99 \\ 
J1807$-$2500B & 0.92  &  1.00  &  1.00 \\  
B1913+16      & 0.68  &  0.92  &  0.94 \\  \hline \hline
\end{tabular}
\label{tb:dns_gammas}
\end{table*}

\begin{figure}
\centerline{\psfig{figure=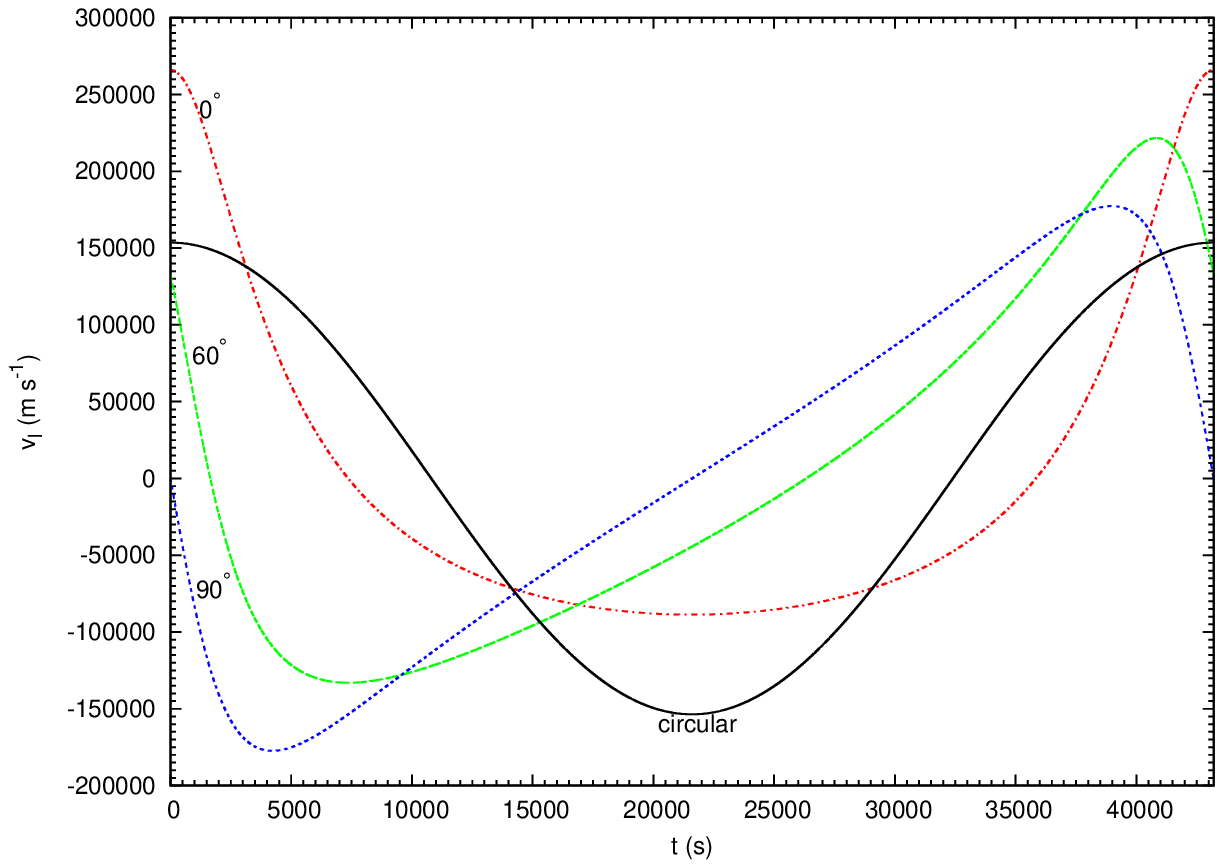,width=8cm,angle=0}}
\caption{Variation of the line-of-sight velocity with time over a complete orbit for a DNS, having $M_{p}=1.35~{M_{\odot}}$, $M_{c}=1.25~{M_{\odot}}$, $P_{o}=0.5$ day, $i = 60^{\circ}$, $T_p = 0$, $e=0.5$ for different $\varpi$, and the line-of-sight velocity for the same binary having zero eccentricity. } \label{fig:radialvelnsns}
\end{figure}

In Fig.~\ref{fig:radialvelnsns} we show the line-of-sight velocity curves for
a hypothetical DNS having $M_{p}=1.35~{M_{\odot}}$,
$M_{c}=1.25~{M_{\odot}}$, $P_{o}=0.5$ day, $i = 60^{\circ}$, $T_p =
0$, $e=0.5$ for different $\varpi$. DNS systems can have such high
eccentricity and short orbital period. One example is the Hulse-Taylor
binary PSR B1913+16, which has $e=0.617$ and $P_{o}=0.323$
day. Comparing Fig.~\ref{fig:radialvelnsns} with
Fig.~\ref{fig:radialvelnswd}, we see that the amplitudes of radial
velocity curves are $\sim 3$ times larger for a DNS in comparison with
a NS-WD binary having the same values of $P_{o}$, $e$, $\sin i$ and
$\varpi$. This fact leads to a reduction in the efficiency factors for
DNSs in comparison with NS-WD binaries.

\begin{figure*}
 \begin{center}
\hskip -2cm \subfigure[$\gamma_{1m}$]{\label{subfig:gamma1_NSNS}\includegraphics[width=0.6\textwidth,angle=0]{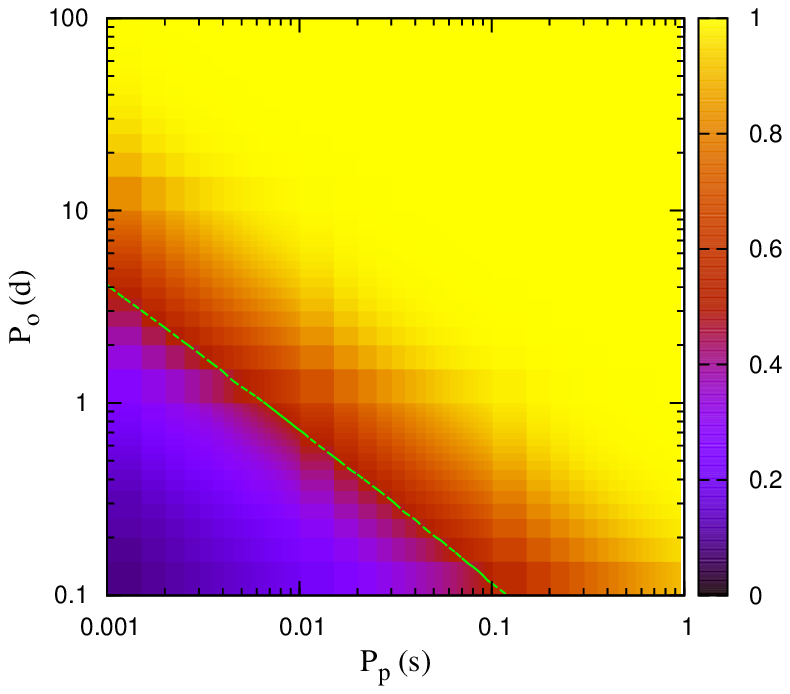}}
\hskip -2cm \subfigure[$\gamma_{2m}$]{\label{subfig:gamma2_NSNS}\includegraphics[width=0.6\textwidth,angle=0]{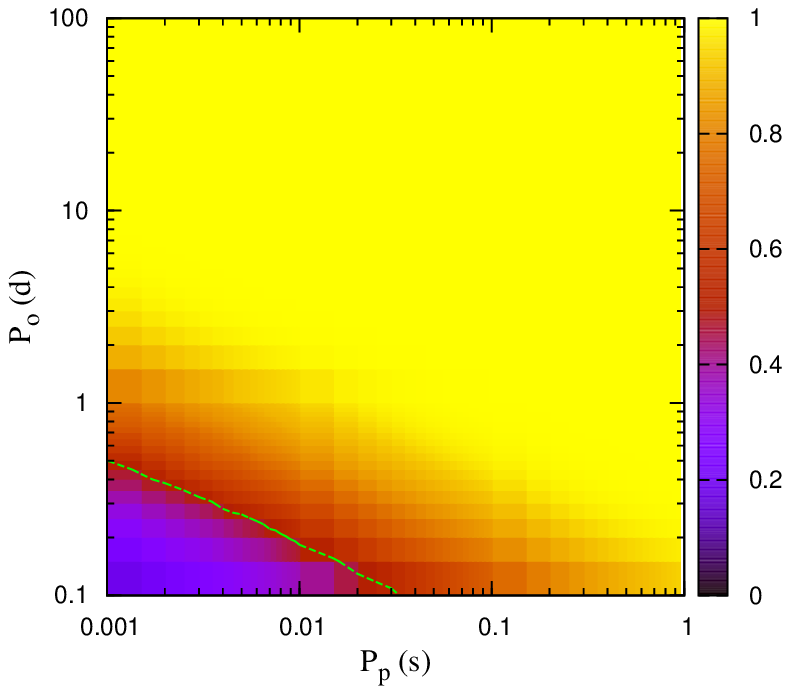}}
\hskip -2cm \subfigure[$\gamma_{3m}$]{\label{subfig:gamma3_NSNS}\includegraphics[width=0.6\textwidth,angle=0]{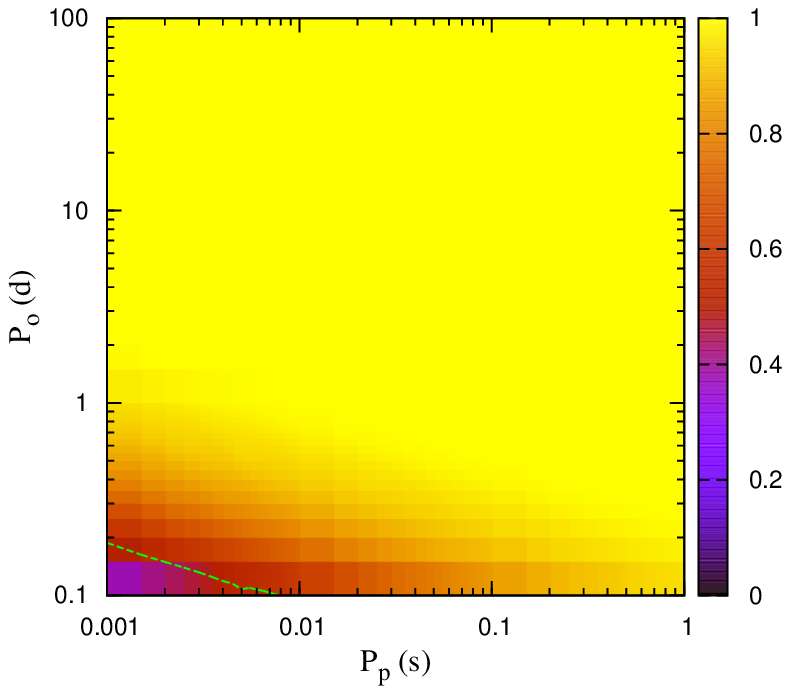}}
 \end{center}
\caption{Variation of $\gamma_{1m}$, $\gamma_{2m}$, and $\gamma_{3m}$ with $P_{o}$ and $P_{s}$ for $M_{p}=1.35~M_{\odot}$,$M_{c}=1.25~M_{\odot}$, $\varpi = 30^{\circ}$, $i= 60^{\circ}$, $e=0.5$, $m=4$, and $T=1000$ s.  }
\label{fig:gammaall_NSNS}
\end{figure*}

\begin{figure*}
 \begin{center}
\hskip -2cm \subfigure[$e=0.1$]{\label{subfig:gamma1ec1_NSNS}\includegraphics[width=0.6\textwidth,angle=0]{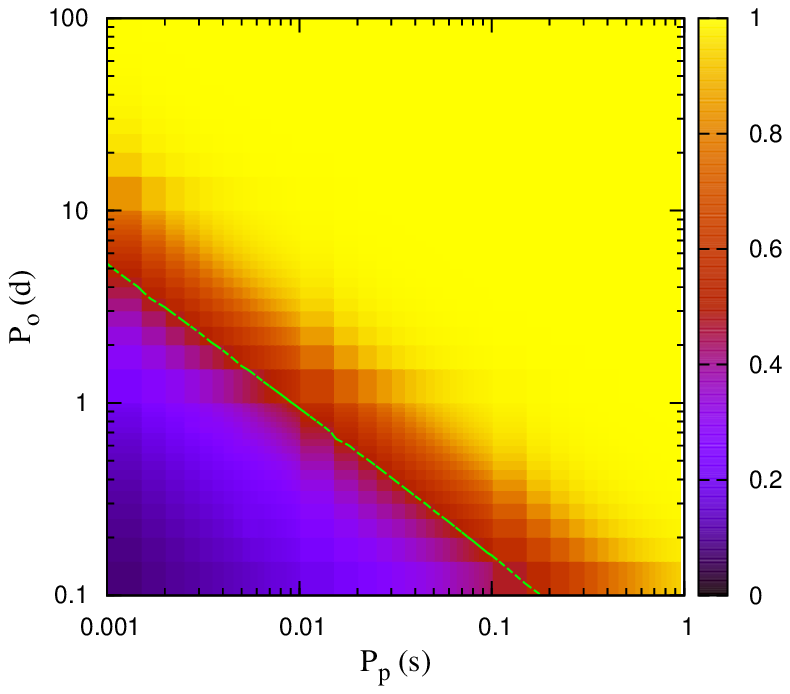}}
\hskip -2cm \subfigure[$e=0.8$]{\label{subfig:gamma1ec8_NSNS}\includegraphics[width=0.6\textwidth,angle=0]{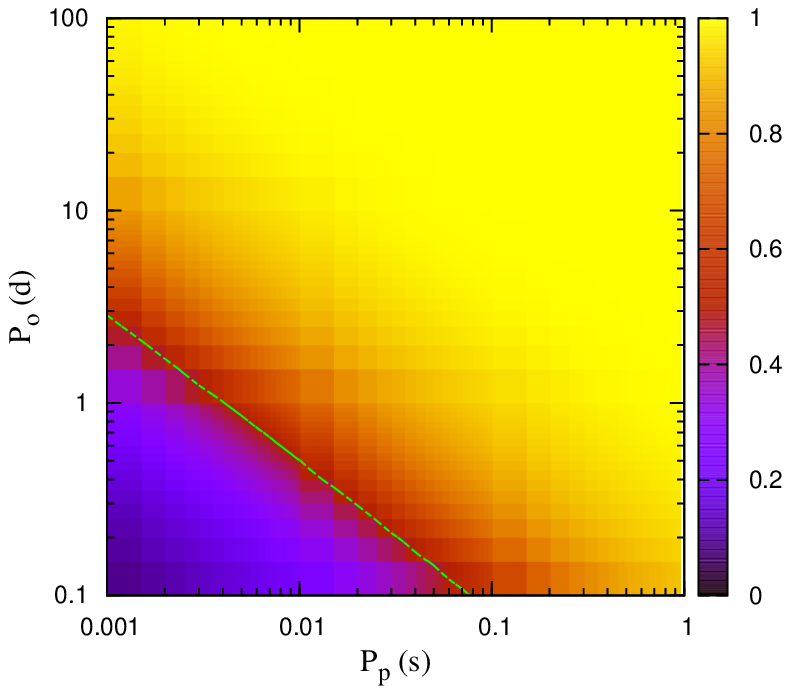}}
 \end{center}
\caption{Variation of $\gamma_{1m}$ with $P_{o}$ and $P_{s}$ for $M_{p}=1.35~M_{\odot}$, $M_{c}=1.25~M_{\odot}$, $\varpi = 30^{\circ}$, $i= 60^{\circ}$, $m=4$, and $T=1000$ s, for $e=0.1$ and $e=0.8$. }
\label{fig:gamma1ec_NSNS}
\end{figure*}

In Fig \ref{fig:gammaall_NSNS}, we show the variations of $\gamma_{1m}$, $\gamma_{2m}$, and $\gamma_{3m}$  with $P_{o}$ and $P_{s}$ for $M_{p}=1.35~M_{\odot}$, $M_{c}=1.25~M_{\odot}$, $\varpi = 30^{\circ}$, $i= 60^{\circ}$, $e=0.5$, $m=4$, and $T=1000$ s. The variations of $\gamma_{1m}$ with $P_{o}$ and $P_{s}$ for $e=0.1$ and $e=0.8$ keeping all other parameters unchanged are shown in Fig \ref{fig:gamma1ec_NSNS}. Again, it is clear that a higher value of eccentricity is favourable for detection, i.e. results a higher value of $\gamma_{1m}$. This fact was also observed by \citet{rzw98}, who studied eccentric systems only for DNS binaries.

%%%%%%%%%%%%%%%%%%%%%%%%%%%%%%%%%%%%%%%%%%%%
%%%%%%%%%%%%%%%%%%%%%%%%%%%%%%%%%%%%%%%%%%%%
%%%%%%%%%%%%%%%%%%%%%%%%%%%%%%%%%%%%%%%%%%%%
%%%%%%%%%%%%%%%%%%%%%%%%%%%%%%%%%%%%%%%%%%%%

\subsection{Neutron Star - Black Hole (NS-BH) Binaries}
\label{subsec:res_nsbh}

Although no neutron star - (stellar mass) black hole (NS-BH) binary is known at
present, there is high possibility of detecting such systems with
forthcoming facilities like SKA \citep{skslcf09}. 

A number of theoretical studies \citep{ppr05, kh09, fl11} exist which
predict probable parameters for these systems. The population synthesis
model of \citet{ppr05} predicts mildly recycled pulsars in short
orbits ($P_{o}$ in the range of $0.004 - 0.8$ days) with small
eccentricities, while the study by \citet{kh09} predicts recycled
pulsars with orbital periods ranging from 0.1 days to $10^5$
days. \citet{fl11} found that recycled pulsars with black hole
companions in highly eccentric orbits with $P_{o}$ between 0.003 and
0.6 days can form due to stellar encounters in dense stellar
environments. In this work, we calculate the efficiency factors for $
0.1 < P_{o} < 100$ days and $ 0.001 < P_{p} < 1$, as we chose for the
cases of NS-WD and NS-NS binaries. We take the mass of the black hole
as 10~$M_{\odot}$, which agrees with the observations \citep{fsc11}.

\begin{figure}
\centerline{\psfig{figure=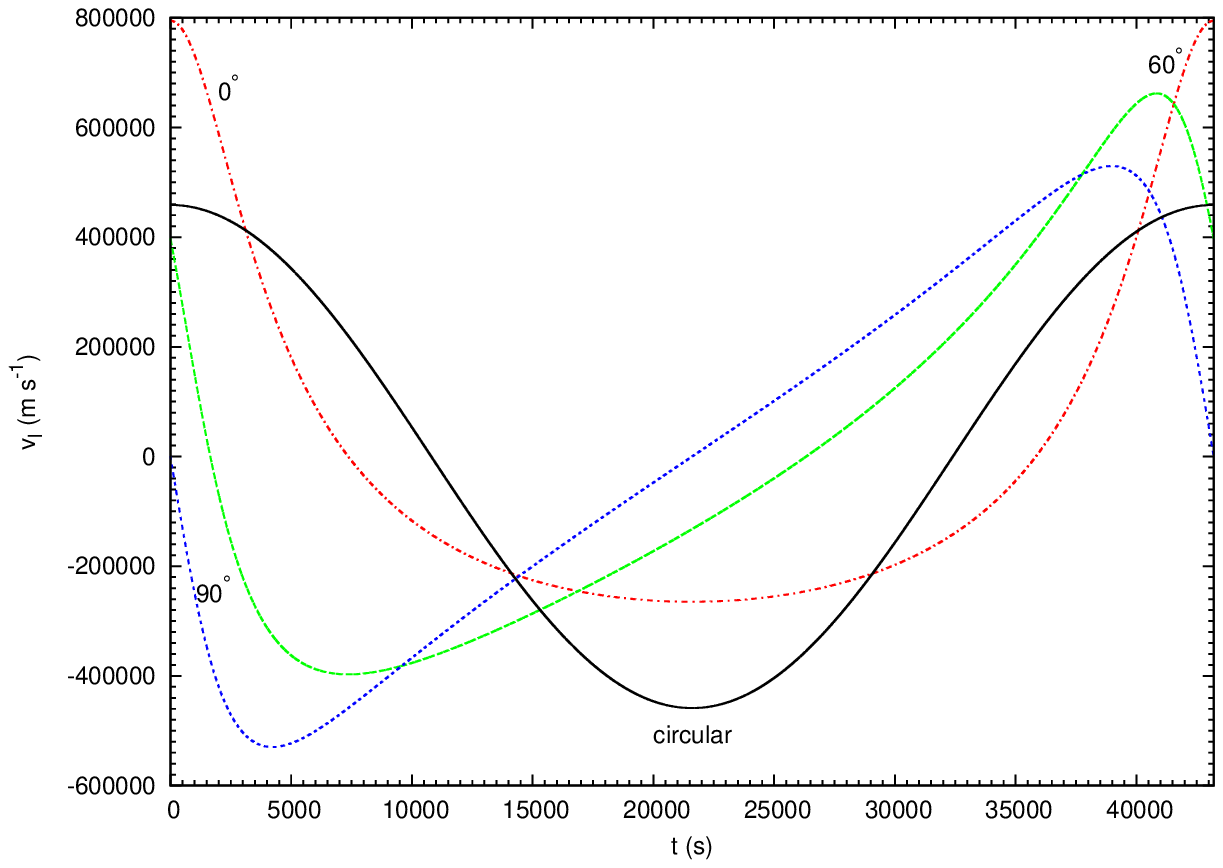,width=8cm,angle=0}}
\caption{Variation of the line-of-sight velocity with time over a complete
  orbit for a binary pulsar with a black hole companion with
  $M_{p}=1.4~{M_{\odot}}$, $M_{c}=10.0~{M_{\odot}}$, $P_{o}=0.5$ day,
  $i = 60^{\circ}$, $T_p = 0$ and $e=0.5$ for different $\varpi$, and
  the line-of-sight velocity for the same binary having zero eccentricity.
} \label{fig:radialvelnsbh}
\end{figure}

In Fig.~\ref{fig:radialvelnsbh}, we show the variation of the radial
velocity with time for a NS-BH binary with $M_{p}=1.4~{M_{\odot}}$,
$M_{c}=10.0~{M_{\odot}}$, $P_{o}=0.5$ day, $i = 60^{\circ}$, $T_p =
0$, $e=0.5$ for different $\varpi$, as well as the circular case
(keeping all other parameters fixed). We see that the amplitudes of the line-of-sight velocity curves are $\sim 3$ times larger for a NS-BH in comparison with a NS-NS binary having the same values of $P_{o}$, $e$,
$\sin i$ and $\varpi$ (or $\sim 9$ times larger for a NS-BH in
comparison with a NS-WD binary). So it is clear that $T = 1000$ s is
too high for this case and will lead to very small efficiency factors
and we decided to use $T = 500$ s.

\begin{figure*}
 \begin{center}
\hskip -2cm \subfigure[$\gamma_{1m}$]{\label{subfig:gamma1_NSBH}\includegraphics[width=0.6\textwidth,angle=0]{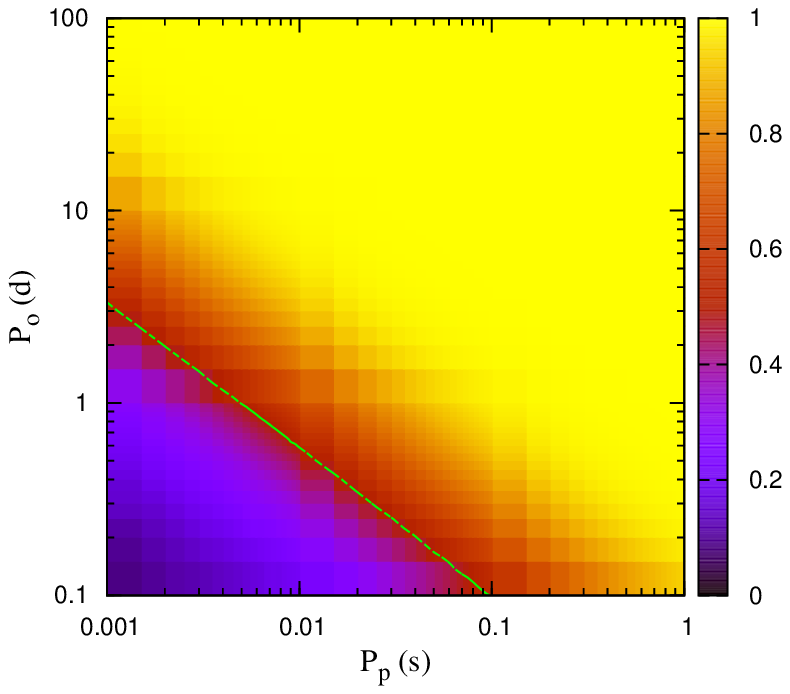}}
\hskip -2cm \subfigure[$\gamma_{2m}$]{\label{subfig:gamma2_NSBH}\includegraphics[width=0.6\textwidth,angle=0]{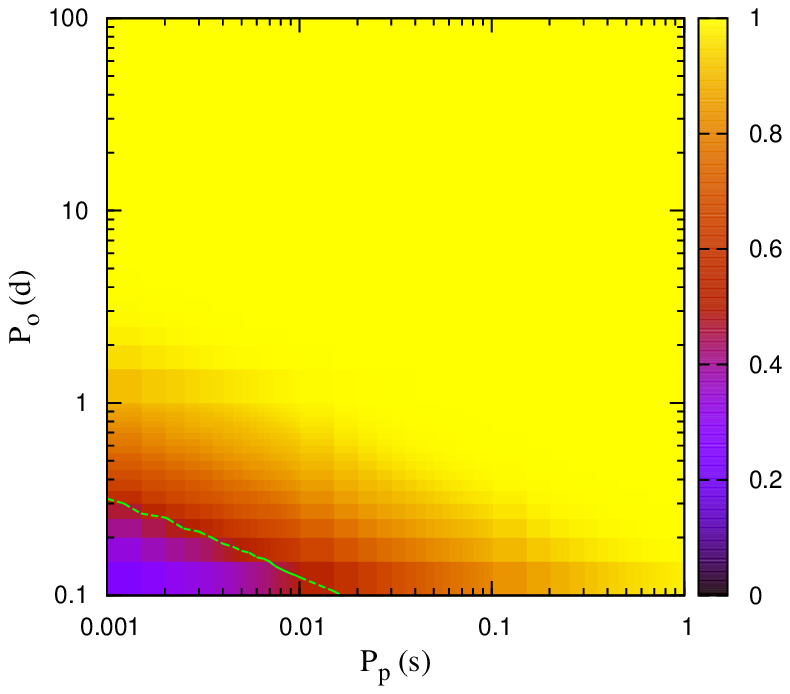}}
\hskip -2cm \subfigure[$\gamma_{3m}$]{\label{subfig:gamma3_NSBH}\includegraphics[width=0.6\textwidth,angle=0]{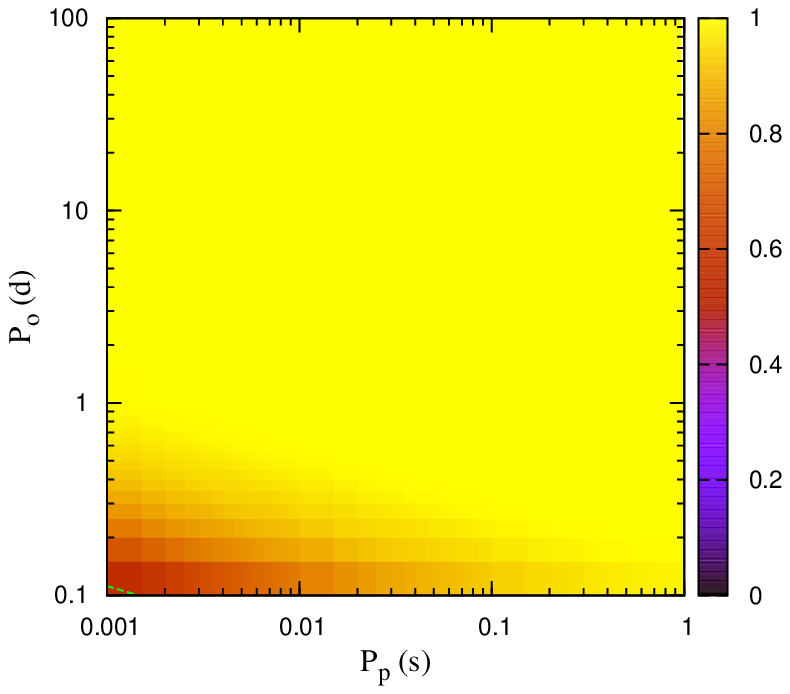}}
 \end{center}
\caption{Variation of $\gamma_{1m}$, $\gamma_{2m}$, and $\gamma_{3m}$ with $P_{o}$ and $P_{s}$ for $M_{p}=1.4~M_{\odot}$, $M_{c}=10.0~M_{\odot}$, $\varpi = 30^{\circ}$, $i= 60^{\circ}$, $e=0.5$, $m=4$, and $T=1000$ s. }
\label{fig:gammaall_NSBH}
\end{figure*}

\begin{figure*}
 \begin{center}
\hskip -2cm \subfigure[$e=0.1$]{\label{subfig:gamma1ec1_NSBH}\includegraphics[width=0.6\textwidth,angle=0]{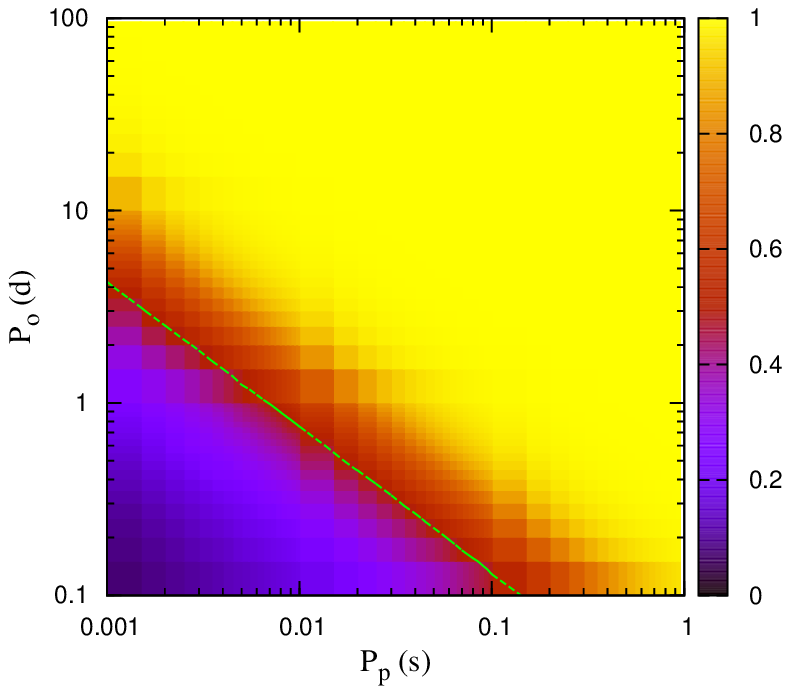}}
\hskip -2cm \subfigure[$e=0.8$]{\label{subfig:gamma1ec8_NSBH}\includegraphics[width=0.6\textwidth,angle=0]{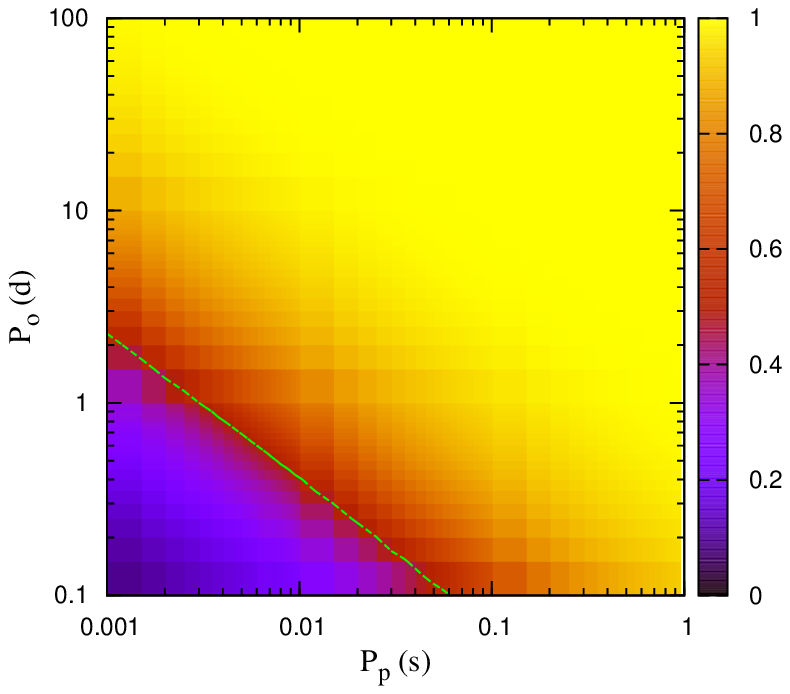}}
 \end{center}
\caption{Variation of $\gamma_{1m}$ with $P_{o}$ and $P_{s}$ for $M_{p}=1.4~M_{\odot}$,
  $M_{c}=10.0~M_{\odot}$, $\varpi = 30^{\circ}$, $i= 60^{\circ}$, $m=4$, and $T=1000$ s, for $e=0.1$ and $e=0.8$. }
\label{fig:gamma1ec_NSBH}
\end{figure*}

In Fig. \ref{fig:gammaall_NSBH}, we show the variations of $\gamma_{1m}$, $\gamma_{2m}$, and $\gamma_{3m}$ with $P_{o}$ and $P_{s}$ for $\varpi = 30^{\circ}$, $i= 60^{\circ}$, $e=0.5$, $m=4$, and $T=1000$ s. The variations of $\gamma_{1m}$ with $P_{o}$ and $P_{s}$ for $e=0.1$ and $e=0.8$ keeping all other parameters unchanged are shown in Fig \ref{fig:gamma1ec_NSBH}. A higher value of eccentricity is favourable for detection, i.e. results a higher value of $\gamma_{1m}$.

%%%%%%%%%%%%%%%%%%%%%%%%%%%%%%%%%%%%%%%%%%%%
%%%%%%%%%%%%%%%%%%%%%%%%%%%%%%%%%%%%%%%%%%%%
%%%%%%%%%%%%%%%%%%%%%%%%%%%%%%%%%%%%%%%%%%%%
%%%%%%%%%%%%%%%%%%%%%%%%%%%%%%%%%%%%%%%%%%%%

\section{Conclusions}
\label{sec:conclu}

We have generalized the earlier work of JK91 to calculate the
signal-to-noise degradation of pulsars in binary orbits with 
arbitrary eccentricity. We have applied the framework to compute
degradation factors for a variety of orbital configurations, and
show that it can be quite substantial. We have also demonstrated how the degradation can be recovered by using acceleration search or 
acceleration-jerk search algorithms. The analysis should prove 
invaluable to a wide variety of population studies since this work
provides, for the first time, accurate accounting for this important
effect. Software to calculate the degradation factors for arbitarary
orbital parameters, harmonic summing and survey integrations is available online at http://psrpop.phys.wvu.edu/binary.

\section*{Acknowledgements}

This work was supported by a Research Challenge Grant to the WVU
Center for Astrophysics by the West Virginia EPSCoR foundation, and
also from the Astronomy and Astrophysics Division of the National
Science Foundation via a grant AST-0907967.

%%%%%%%%%%%%%%%%%%%%%%%%%%%

\end{document}